\documentclass[traditabstract]{aa} 
\usepackage{graphicx}
\usepackage{txfonts}

\def\sqdeg {\,{\rm deg^2}} 
\def\mic {\, \mu{\rm m}} 
\def\Ks {{\rm K_s}} 
\def\mm {\, {\rm mm}} 
\def\newt {\, {\rm N}} 
\def\degc {\, ^{\circ}{\rm C}} 

\def\simlt{\mathrel{\lower0.6ex\hbox{$\buildrel {\textstyle <} 
  \over {\scriptstyle \sim}$}}}
\def\simgt{\mathrel{\lower0.6ex\hbox{$\buildrel {\textstyle >}
 \over {\scriptstyle \sim}$}}}

\def\newtwo{ } 

\begin{document}
   \title{The Visible and Infrared Survey Telescope for Astronomy (VISTA):
 Design, Technical Overview and Performance}


   \author{Will Sutherland\inst{1}, 
 Jim Emerson\inst{1}, Gavin Dalton\inst{2,3}, 
 Eli Atad-Ettedgui\inst{4,5}, Steven Beard\inst{4}, 
  Richard Bennett\inst{4}, Naidu Bezawada\inst{4}, 
 Andrew Born\inst{4},  Martin Caldwell\inst{2}, 
 Paul Clark\inst{6}, 
 Simon Craig\inst{7}, David Henry\inst{4},  
 Paul Jeffers\inst{7}, Bryan Little\inst{4},  Alistair McPherson\inst{8}, 
 John Murray\inst{4}, Malcolm Stewart\inst{9}, 
 Brian Stobie\inst{4}, David Terrett\inst{2}, 
 Kim Ward\inst{2}, Martin Whalley\inst{2}, Guy Woodhouse\inst{2}. 
 } 

 \institute{School of Physics and Astronomy, 
   Queen Mary University of London, Mile End Rd, London E1 4NS, UK. \\
              \email{w.j.sutherland@qmul.ac.uk}
  \and
   RAL Space, Harwell Oxford, Didcot, Oxfordshire OX11 0QX, UK.  
 \and 
  Astrophysics, University of Oxford, Keble Road, Oxford OX1 3RH, UK. 
 \and
   UK Astronomy Technology Centre, Royal Observatory, Blackford Hill, 
    Edinburgh EH9 3HJ, UK. 
 \and 
 Senior Optical Consultant, 9 Abercorn Road, Edinburgh EH8 7DD, UK. 
 \and 
  Centre for Astronomical Instrumentation, 
  University of Durham, South Road, Durham DH1 3LE, UK. 
 \and 
  National Solar Observatory, NSO/DKIST, 
  950 N. Cherry Avenue, Tucson, AZ 85719, USA. 
  \and
  SKA Organisation, Jodrell Bank Observatory, Lower Withington,
 Macclesfield, Cheshire SK11 9DL, UK.  
 \and 
  Solaire Systems, 55/10 Bath Street, Edinburgh EH15 1HE, UK. 
  } 
             
 \date{A\&A Received 12 September 2014 / Accepted 15 December 2014 }
 \authorrunning{W. Sutherland et al.} 
 \titlerunning{VISTA: Technical Overview and Performance} 

 
 \abstract{ 
 The Visible and Infrared Survey Telescope for Astronomy (VISTA) 
 is the 4-metre wide-field survey telescope at ESO's Paranal Observatory, 
 equipped with the world's largest near-infrared imaging camera 
 (VISTA IR Camera, VIRCAM), 
 with 1.65 degree diameter field of view,  and
  67 Mpixels giving 0.6 $\sqdeg$ active pixel area,  
  operating at wavelengths $0.8 - 2.3 \mic$.  
 We provide a short history of the project, and 
  an overview of the technical details of the full system including 
 the optical design, mirrors, telescope structure, IR camera, active optics, 
  enclosure and software.  
   The system includes 
  several innovative design features such as the 
  $f/1$ primary mirror, the {\newtwo dichroic} cold-baffle camera design and 
  the sophisticated wavefront sensing system delivering   
   closed-loop 5-axis alignment of the secondary mirror. 
  We conclude with a summary of the 
   delivered performance, and a short overview of the six ESO public
  surveys in progress on VISTA.   
 } 

   \maketitle
%

\section{Introduction}
\label{sec-intro} 

Wide-field imaging surveys have long formed a cornerstone of
 observational astronomy, from the photographic Schmidt telescope surveys
 from Palomar, the UK Schmidt Telescope and 
  ESO in the 1950--1980 era. After the
 advent of large-format CCDs and near-IR detectors during the 1990s,
 these were followed by a 
 number of major multi-colour digital sky surveys,  
 notably the Sloan Digital Sky Survey
 (SDSS; Gunn et al \cite{gunn}, Abazajian et al \cite{sloan-dr7}), 
the 2 Micron All-Sky Survey (2MASS; Skrutskie et al \cite{tmass}), 
 the CFHT Legacy Survey (CFHTLS; Cuillandre et al \cite{cfhtls}),  
 and the UK Infrared Deep Sky Survey (UKIDSS; Lawrence et al \cite{ukidss}).  
 As well as forming a fundamental 
 legacy resource (notably as an atlas 
  for identifications of sources discovered 
 at many other wavelengths, from radio and sub-mm to 
 X-rays and gamma rays), 
  these surveys have led to a very wide
   range of new discoveries covering most
  areas and scales of observational astronomy, 
  ranging from asteroids, brown dwarfs, 
   Galactic structure, new Milky Way satellite galaxies, 
  through external galaxies and clusters, 
  out to large-scale structure, weak lensing and the highest 
  redshift quasars. 

In late 1998, a new science funding opportunity 
 was provided by the UK Joint Infrastructure Fund; 
 a consortium of 18 UK universities (see Acknowledgements) 
  put together a successful
 proposal to build a new 4-metre class 
  wide-field survey telescope, sited in the Southern hemisphere and  
  mainly dedicated to multicolour imaging surveys; this was named 
 the Visible and Infrared Survey Telescope for Astronomy (VISTA). 
 Since 2009, the VISTA telescope and its near-infrared
  camera (VIRCAM) have been in science operations
  at ESO's Paranal Observatory: the product of 4 metre aperture 
 and $0.6 \sqdeg$ on-pixel field of view makes VISTA the world's fastest 
  near-infrared survey system,  and it seems likely to
 retain this advantage until the launch of 
 a dedicated space mission such as {\em Euclid} or {\em WFIRST} 
 in around 2020. 

\begin{figure*}[htb!]
   \centering
\includegraphics[width=15cm]{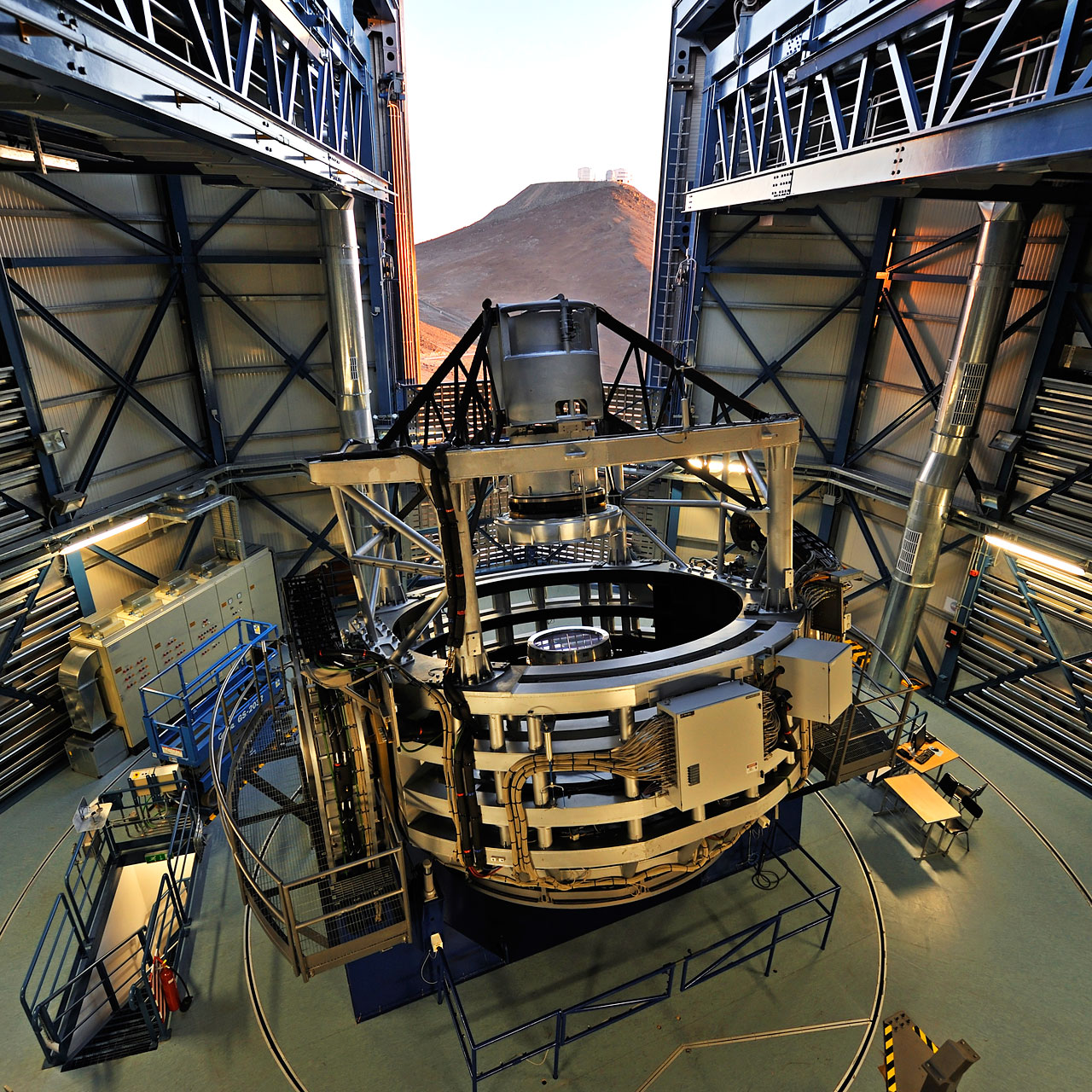}
 \caption{VISTA telescope at sunset, 
 with the main Paranal summit and VLTs in the background. 
 The VIRCAM vacuum window
  is visible in the centre of the tube. On the telescope top-end, 
  the M2 hexapod
 is behind the top ring; the M2 Cell (black) below it, and the M2 Baffle 
 is the metallic annulus. (Photo credit: G.Hudepohl/ESO) } 
\label{FigTel} %
\end{figure*}

 This paper provides an overview of the development, 
 technical details and on-sky performance of the VISTA telescope
  and VIRCAM. The intention is to provide
 an intermediate level of detail across the whole system, 
  plus references to more detailed papers for each subsystem.  
 The remainder of the paper is divided into sections  
  as follows: 
 in Sect.~\ref{sec-over} we provide a short outline of
  the full system, to place in context the
  more detailed later sections.    
 In Sect.~\ref{sec-hist} we outline the early history of the
  project and rationale of the basic design choices; 
 in Sect.~\ref{sec-des} we provide an overview of the system optical
  design and overall layout; Sect.~\ref{sec-mirrs}
 describes the two mirrors and their support systems; 
 Sect.~\ref{sec-tel} describes the telescope 
  structure and axes rotation systems; Sect.~\ref{sec-cam} describes the
  infrared camera; Sect.~\ref{sec-actopt} describes the 
  active optics control; 
 Sect.~\ref{sec-encl} describes the enclosure and ancillary
  equipment; Sect.~\ref{sec-soft} describes computer control
 and software; 
  Sect.~\ref{sec-commiss} summarises the commissioning; 
 Sect.~\ref{sec-data} summarises observation control, 
  data processing and archiving; 
 and Sect.~\ref{sec-perf} gives a short summary of
  the operational performance. 
 These sections are largely self-contained, so the reader
 interested in particular aspects is advised to read Sect.~\ref{sec-over}
 then skip to the specific section(s) of interest. 

\section{System Overview} 
\label{sec-over} 

The VISTA telescope (Fig.~\ref{FigTel}) 
 is located at ESO's Cerro Paranal Observatory 
 in northern Chile, 
 at latitude $24^\circ \, 36^\prime \, 57^{\prime\prime}$ South, 
 longitude $70^\circ\, 23^\prime \, 51^{\prime\prime}$ West; this is
 approximately 120\,km south of Antofagasta city, and 12\,km from
  the Pacific coast.  
 Locally, VISTA is sited on a subsidiary summit of elevation 2518\,m,  
  approximately 1.5\,km NNE from the 
  main Paranal summit which hosts the four VLTs, the 
 VLT Interferometer and the VLT Survey Telescope;  
  thus VISTA shares the same outstanding weather conditions.  
 Though VISTA is at approximately 100\,m lower elevation than the main 
  summit, it is rarely downwind of the main summit. 
 Clearly, the Southern hemisphere site is very important for
 complementarity with next-generation major projects including ALMA, 
  SKA Pathfinders and E-ELT, since (apart from 2MASS) 
  the major wide-area optical and near-IR 
  surveys since 2000 (notably SDSS, CFHT Legacy Survey, UKIDSS) are all 
  concentrated in the Northern hemisphere. 

 During routine observing, VISTA is operated entirely in queue-scheduled mode, 
  controlled remotely by a single telescope operator 
  in the main VLT control room; evening startup and
 morning shutdown procedures are done by an operator 
 adjacent to the telescope for safety reasons. 

\begin{figure}[htb!]
   \centering
\includegraphics[width=9cm]{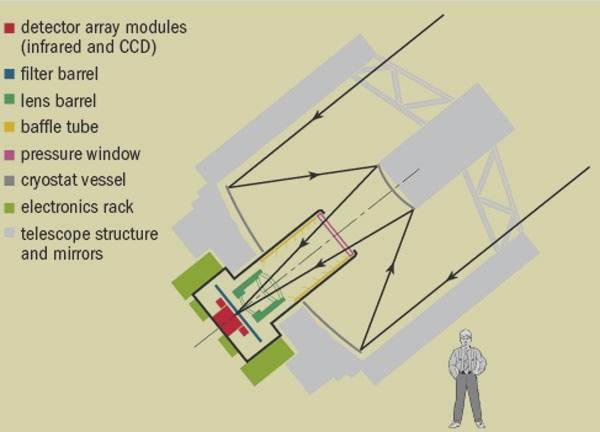}
 \caption{A schematic view of the general layout of VISTA optical
  components; this shows the two mirrors, the VIRCAM
 entrance window and lenses, and the main 
  components of VIRCAM.} 
\label{Fig-Schem} %
\end{figure}

 The telescope optics (Sect.~\ref{sec-des}) 
  use a very fast two-mirror quasi-Ritchey-Chretien design.  
 The 4.1 m diameter primary mirror is a 
 hyperboloid of $f/1.0$ focal ratio; 
 together with a 1.24 m convex hyperboloid secondary mirror, 
 this produces an $f/3.25$ Cassegrain focus located behind the M1 cell,
  $\approx 1.17$\,m below the M1 pole (Fig.~\ref{Fig-Schem}). 
 The Cassegrain is the only
 focal station, so VISTA carries one large instrument at any time;  
  currently the IR Camera is the only instrument available for VISTA.   
 The telescope is designed for occasional instrument interchange, and
  a wide-field 2400-fibre multi-object spectrograph 
 (4MOST; de Jong et al \cite{f-most}) is approved by ESO for 
  installation after 2019. 
 The telescope uses active (not adaptive) optics; 
  the primary mirror has an 84-point active axial support
  system, while the secondary mirror is mounted on
 a hexapod for 5-axis position control. Physical details of the two
  mirrors and their support systems are in Sect.~\ref{sec-mirrs}. 

 The telescope structure (Sect.~\ref{sec-tel}) 
  is a conventional though very compact Alt-azimuth mount, 
  constructed mainly of steel, with a 
  moving mass of 90 tonnes and sweep radius of 4.6\,m. 
 The telescope uses rolling-element bearings 
   on all three rotation axes, 
  and direct position encoding using 
  Heidenhain optical tape encoders with four read heads 
  per axis. The telescope axes are driven via counter-torqued gearboxes 
  for the Azimuth axis and Cassegrain rotator, 
  and two direct-drive motors for the
  Altitude axis.  

 The very large VISTA IR Camera (VIRCAM, Sect.~\ref{sec-cam}) is mounted on 
  the Cassegrain rotator on the back of the primary mirror cell; 
 the camera mass is 2.9\,tonnes including 800\,kg cold, 
  and the cryostat length is 2.8\,m. 
 From front to back, the camera includes 
  a 95\,cm diameter vacuum window, 
  a long cold--baffle tube
  to minimise thermal background on the detectors, 
 a lens barrel with three Infrasil field-corrector lenses, 
 an 8-position filter wheel of 1.37\,m diameter, 
 and sixteen Raytheon VIRGO $2048^2$ 
  HgCdTe near-infrared detectors, giving a mean
 pixel scale of 0.339 arcsec. The detectors are arranged
 in a sparse-filled $4\times 4$ rectangular grid 
  within the $1.65$ degree (350 mm) diameter
 field of view, and provide an active field of $0.60\sqdeg$ 
 on pixels.  

 The VIRCAM also includes two fixed autoguiders, two fixed low-order wavefront
 sensors (using CCDs), and movable beamsplitters feeding the science
 detectors for high-order wavefront sensing, to
 provide guiding and active optics corrections respectively; details
  are provided in Sect.~\ref{sec-actopt}.  

 The telescope is housed in a 19\,m diameter enclosure (details
  in Sect.~\ref{sec-encl}), which 
  follows standard modern design practice to minimise local
  seeing effects. Features include
  a powerful air-cooling system to maintain the interior at nighttime
  temperature during the day, active cooling of
  all electronic boxes in the dome,  six large ventilation doors 
  and a movable windscreen to optimise airflow during observing,
  and a movable moonscreen for reduction of stray light.  
  A single-storey auxiliary building, adjoining 
  the main enclosure on the South side, 
 houses infrastructure facilities including the mirror coating plant, 
  electrical distribution units and glycol pumps.  


 The system is designed mainly for efficient surveying of 
  large areas of sky: clearly the main design drivers for this are
 the very wide field of view, moderately large aperture, and the
  high QE of the detectors; other important contributing factors
  are the high fraction $\ge 75\%$ of
  observing time dedicated to large-scale survey programs, 
 and the minimisation of times for necessary observing overheads, e.g. 
  detector readout and telescope 
  jitter movements.  
 An overview of the on-sky performance is given in Sect.~\ref{sec-perf}.

\section{Basic requirements and design selection} 
\label{sec-hist} 
\subsection{Top-level specifications} 
 At the start of the project `Phase A' design in April 2000,  
  the basic requirements were as follows. The telescope aperture was set to 
  approximately 4 metres for budget and schedule 
 reasons, availability of an existing mirror blank, 
  and also because achieving a 
 wide field of view becomes progressively more challenging on larger
 telescopes. The initial baseline was to accommodate two cameras: 
  a $0.3 \sqdeg$ 36 Mpixel (9 detector) near-infrared camera,  
   and a $2 \sqdeg$ 450 Mpixel visible camera,  
  both with image quality commensurate with seeing-limited images
  at a top-class ground-based site. 
 A key goal was to permit an upgrade path for the IR Camera 
  to 67 Mpix and 16 detectors: 
 this upgrade subsequently came to reality, using additional
  funding provided by the UK, as part of the in-kind 
 contributions for UK entry into ESO membership. 

 The site was required to be in the Southern hemisphere, since
 all the leading wide-field imaging systems then operating (SDSS) or under 
 construction (CFHT-Megacam, Subaru-SuprimeCam, UKIRT-WFCAM) 
 were sited in the Northern hemisphere. At the time of VISTA commissioning
  the only wide-area optical/near-IR Southern surveys were 
  photographic Schmidt plates and the 2MASS survey; subsequently
  the Skymapper (Keller et al \cite{skymapper}), 
  VST (Cappacioli \& Schipani \cite{vst}) 
  and DECam (Flaugher et al \cite{decam}) 
 instruments have begun Southern surveys at visible wavelengths which
 are important complements to VISTA.   
 The site was chosen to be
  Cerro Paranal in early 2000, and this proved later to be an important 
   step in the UK's subsequent joining of ESO in 2001--2.  

 The original proposal included two cameras, infrared and visible, 
  with the infrared camera taking precedence in tradeoffs.  
  The proposed visible camera  was eventually not built for a combination of
  several reasons: it was initially postponed due to funding constraints; and
  the UK's entry to ESO included access to the 2.6\,m VST survey 
  telescope, predicted to be operational earlier than VISTA.    
 However, capability for a visible
  camera with field diameter $\ge 2.1$ degrees 
 was fully preserved in the telescope design.   
 
 The general concept was that the telescope was designed 
  exclusively as the front-end to
 the IR and visible cameras, and thus general purpose
  facilities such as additional foci, 
  night-time instrument changes or mid-infrared operations 
  were given zero weight to constrain cost.  
 Spectroscopic capability was also not required, though the final 
  telescope design turned out to be 
  very suitable for a fibre-fed multi-object spectrograph, in 
  particular 4MOST (see Sect.~\ref{sec-perf}). 
 
 One key design parameter was the pixel scale in arcseconds: 
  early in the Phase A study, the Science Committee 
 converged on required pixel scales close to 0.33 arcsec for infrared imaging,
  and 0.25 arcsec for visible imaging, with 10 percent tolerance.   
 The rationale for larger near-IR pixels is due to cost: 
  although atmospheric seeing is slightly better in the near-IR, the
   cost per pixel of near-IR detectors is 
 $\sim 20\times$ higher than visible CCDs due to the
 much greater manufacturing complexity and smaller commercial market. 
 Therefore, whereas 
  for a visible imager it is generally affordable to tile the 
  entire correctable field with CCD detectors, this is not true 
  in the near-IR; so the tradeoff between sky coverage and 
 sampling favours coarser sampling in the near-IR than the visible.  
 At that time, standard pixel sizes were $13-15 \mic$ 
 for CCDs and $18-20 \mic$ for near-IR detectors, and 
 there was a large cost penalty for non-standard pixel sizes: 
 thus, a physical plate scale near $60 \mic$/arcsec was required for both
  cameras, which translates to a final focal length near 12 metres.  

 Another key design requirement was for the infrared imager to operate
  efficiently from Y to $\Ks$ bands, 
 i.e. long wavelength cutoff $\ge 2.3 \mic$.\footnote{
 We did not consider operation at 
 $\lambda > 2.5 \mic$, since the {\em Spitzer} 
 spacecraft was then in an advanced stage of development, and 
 at $\lambda > 3 \mic$ the foreground advantage in space is 
 overwhelmingly large.
  This has proved the correct decision given the excellent 
   performance of {\em Spitzer} and more recently the {\em WISE} mission.} 
 Operating at $\Ks$ band does
 add significant challenges, since a room-temperature
 black surface has $\sim 10\times$ the surface brightness of
 the night sky at $\Ks$ band;  thus, the detectors must be 
 blocked from viewing any significant solid angle of warm
 high-emissivity surfaces.  

 Also, an IR camera requires at least a transmissive
 vacuum window, and typically several transmissive corrector 
  elements. All standard optical glasses 
 absorb strongly beyond wavelength $\lambda > 2 \mic$, 
 so more exotic materials are required: as the 
  required diameter grows beyond 300\,mm, 
 the list of available IR-transmitting materials rapidly shrinks, 
 leading to substantial challenges keeping chromatic
 aberrations acceptably small with a large physical focal plane.  
 This challenge was overcome using a novel cold-baffle solution,
  as described later in Sect.~\ref{sec-cam}.  

\subsection{Design shortlist} 

We here outline some potential concepts which were studied
  during the early Phase-A studies, before downselecting to
 the final design with $f/1.0$ primary mirror and $f/3.25$ Cassegrain
 focus.   

 The initial JIF proposal was for a 4\,m $f/2$ primary mirror
 with a flipping top-end and two cameras: a visible camera with corrector
 lenses using the prime focus, and an $f/6$ secondary mirror
 feeding a re-imaging near-IR camera at Cassegrain focus 
 (similar to a larger-diameter
  version of UKIRT WFCAM; Casali et al \cite{wfcam}). 
 During the Phase-A study, this evolved slightly 
  to $\approx f/2.5$ primary, to
  provide the necessary prime-focus scale for the visible camera
  with a system $f$-ratio $\approx f/2.8$ after the wide-field corrector.   
 After some investigation, the
  flipping top-end concept was discarded as unreasonably massive, 
  and a variant was studied with two interchangeable
 top-end rings, carrying respectively the visible camera and the 
  IR secondary mirror. This concept was feasible, but required a 
   fairly tall telescope structure and large dome, with geometry
  similar to the William Herschel Telescope.  
 Also, the WFCAM-like layout for the IR camera 
  requires the detector package to be mounted downward-facing, 
  in the beam above the cryogenic tertiary mirror. 
  The resulting tradeoff between 
  tertiary size and central obstruction fraction would make it impractical to
  expand the field of view beyond $\sim 1.2$ degree diameter, so this
  route could not reasonably accommodate an upgrade of 
  the near-IR camera to the goal of 67 Mpixels.  
 
 Refractive IR camera designs such as CFHT-WIRCAM (Puget et al \cite{wircam})
  and KPNO-NEWFIRM 
 (Probst et al \cite{newfirm}) are more compact, 
   but generally require exotic lens materials 
  so again do not readily scale up to fields $ > 1$ degree. 

 In parallel with the $f/2.5$ baseline, other concepts were investigated: two 
 briefly, and one in more detail which was finally adopted in 
 preference to the original. 
 A solution using 2 Nasmyth foci was initially appealing since
 the planned visible-IR instrument change would be a simple
 rotation of the tertiary mirror; however, this concept   
 required the focal planes to be well outside the 
  Altitude bearings, in order to avoid pre-focal corrector
  lenses obstructing the light-path above the primary. 
 This in turn required unreasonably large M2 and M3 mirrors, 
 and very large holes through the altitude bearings, 
  so the Nasmyth option was rejected.

Three-mirror telescope designs have the
 advantage of excellent image quality over an extremely wide field of view, 
 (e.g. 3.5 degrees diameter for the case of the future LSST; Abell et al 
  \cite{lsst}) 
 but there are several penalties: 
 they are significantly more costly for given aperture 
  than 2-mirror systems, have a large central obstruction, and
 would have great difficulty accomodating a large and massive 
 infrared camera near the top-end. Therefore, 3-mirror 
 designs were also rejected.   


In parallel with the original concept, another more compact 
  telescope solution was studied in detail, with an 
  $f/1.0$ primary mirror and a single $f/3.25$ Cassegrain 
  focus, feeding interchangeable IR and visible cameras
 alternately.
(We note here that solutions with $\sim f/1.5$ primary 
 are simple, but could not meet our requirements;  
  the prime focus has too short focal length for an imager, 
  while delivering a $\sim f/3$ Cassegrain focus behind the M1 
  would require an unreasonably large secondary mirror).  

 Compared to the baseline $f/2.5$ primary, the $f/1.0$ solution offered
  lower total cost, reducing moving mass from 250 tonnes
 to 90 tonnes (Craig et al \cite{craig}) and reducing
 enclosure size,  and enabled a significantly wider 
  near-IR field of view permitting an upgrade path to 16 detectors; 
  so the $f/1.0$ design was selected at a downselect review 
  mid-way through the Phase A study, and developed
  thereafter.  
 The final optical design is described in the next section 
 (Table~\ref{tab:pars}). 

\begin{table*} 
\caption{VISTA: main system parameters} 
\label{tab:pars} 
\centering
\begin{tabular}{l l } 
\hline 
  Parameter &  Value  \\ 
\hline 
 Site            &  Paranal Observatory, Chile \\ 
 Primary mirror diameter  & 4.10 \, m  \\ 
 Secondary mirror diameter & 1.241 \, m \\ 
 Focus    &   Cassegrain only \\ 
 System focal length (with IR corrector) & 12.072 \, m \\ 
 M1 - M2 spacing   &   2725.7 \,mm  \\ 
 Mount type  & Alt-azimuth  \\ 
 Moving mass   & 90 tonnes \\ 
 VIRCAM wavelength range &   $ 0.8 - 2.3 \mic$ \\ 
 Field of view (VIRCAM)  & 1.65 degree (diameter) \\ 
 Field on IR pixels & 0.6 $\sqdeg$ \\ 
 Mean pixel scale & 0.339 arcsec \\ 
 VIRCAM science detectors & Sixteen Raytheon-VIRGO HgCdTe, $2048^2$ format \\ 
 VIRCAM guide/wavefront detectors & Six E2V CCD 42-40, $2048^2$ format \\ 
 Filter set (current)   &  Z,Y,J,H,$\Ks$, two narrowbands, one dark.  \\ 
\hline 

\end{tabular} 
\end{table*} 

\begin{table*}  
\caption{VISTA mirror parameters (as-built values) } 
\label{tab:mirrs} 
\centering
\begin{tabular}{l c c c c} 
\hline 
  Mirror  & Physical diameter & Clear aperture & 
    Radius of curvature  & Conic constant \\ 
          & (mm)  &  (mm)    & (mm)  & \\ 
\hline 
 M1  & 4100    & 3960    & 8094.2  &  -1.12979   \\ 
 M2  & 1241.5   &  1240.5    & 4018.9  & -5.5494    \\ 
\hline 

\end{tabular} 
\end{table*} 

\section{Optics Overview} 
\label{sec-des} 

\subsection{Optical design} 
 As described above, 
 the Phase A study selected a novel overall design as follows: 
 the VISTA telescope is a quasi-Ritchey-Chretien 2-mirror telescope 
 with a very fast $f/1.0$ primary mirror, and a moderately large
  1.24\,m secondary mirror giving a single $f/3.25$ Cassegrain 
  focal station (Fig.~\ref{Fig-Schem}); the
  infrared camera VIRCAM (and the conceptual visible camera) 
  uses this direct Cassegrain focus,  
 with any VISTA instrument including its own wavelength-specific 
  field corrector lenses. 

 This leads to several unusual features: firstly, the $f/1.0$ 
 primary mirror (hereafter M1) 
 results in a very compact telescope with
   a separation of only 2.725\,m from M1 to M2.  Also, the
 infrared camera does not use re-imaging or a cold stop: 
 the 1.24\,m secondary mirror is undersized to form the aperture stop,
 so each detector pixel views a 3.70\,m off-centre circle 
  on the primary mirror; the envelope of these is a 3.95\,m circle,
  so no pixel views warm structure outside the primary mirror.    
 The near-IR detectors are located at the corrected telescope focus 
 with no re-imaging optics, and
 the IR camera includes a long ``cold baffle'' cylinder 
 extending 2.2\,m above the focal plane to minimise the 
  detectors' view of warm surfaces.
 (In summary, the IR camera layout is similar to a 
    conventional visible Cassegrain imager,  
  but with the cryostat greatly expanded to 
  enclose the filter changer, the corrector lens barrel and 
   the primary baffle tube).   

 The telescope and IR Camera were designed as a single optical system;  
  the conic constants of both mirrors were optimised
 jointly with the VIRCAM corrector lens surfaces, to optimise image quality
 averaged over the full 1.65 degree diameter field.  
 The axial thicknesses of the IR Camera window, filters and lenses 
  were included in this
  optimisation process, but held fixed for mechanical reasons; 
  lens spacings and curvatures were allowed to vary.  
 This leads to mirror conic constants which are slightly
  different from an exact Ritchey-Chretien system; mirror 
 parameters are given in Table~\ref{tab:mirrs}.   
 (The visible camera was then separately optimised with these
   mirror parameters as fixed inputs).

 Dispensing with re-imaging means that the IR Camera
  corrector lenses only need to correct the off-axis aberrations
  of the telescope: thus they have weak power, 
  so a single material (Heraeus Infrasil low-OH fused silica) 
 can be used for the vacuum window and all three corrector lenses.  
 Unlike most IR lens materials, Infrasil is robust, 
  highly resistant to thermal shock, very homogeneous, and is 
   available in large diameters.  

 Thus, the light path is as follows: after reflection from M1
 and M2, the incoming converging beam 
  is transmitted through the flat Infrasil cryostat window, 
  then through three Infrasil corrector 
 lenses, then the passband filter, then 
  finally reaches focus at the IR detector plane.   

 The 3-lens corrector serves three main functions:
 \begin{enumerate} 
 \item To correct the off-axis aberrations (mainly astigmatism)
    from the 2-mirror telescope. 
 \item To flatten the focal surface to allow a flat array of detectors.  
 \item To correct for chromatic aberrations induced by the (flat)
    vacuum window and filter.  
 \end{enumerate} 
  Five of the six lens surfaces are spherical, with the 
  one aspheric surface on the upper (concave)
  surface of lens 3.   The detector plane is flat (allowing
 slight field curvature was explored, but gave negligible improvement 
 in image quality).  The design delivers high throughput 
 and excellent image quality over a 350\,mm, $1.65$ degree 
 diameter flat field, without the use of cryogenic mirrors 
  or exotic lens materials. 

 The image quality {\em on paper} (including diffraction 
  with the 1.63\,m diameter M2 baffle, but assuming perfect 
 optical surfaces and perfect alignment) has a 50\% encircled-energy
 diameter (EED) of $\le 0.36$ arcsec 
 and an 80\% EED of $\le 0.68$ arcsec across 
 the entire field of view at Y to $\Ks$ bands.  
  Mean values across the field 
  are approximately $0.27$ arcsec (50\% EED) and 0.45 arcsec (80\% EED) 
 with weak passband dependence, 
  leaving a reasonable margin for real-world degradation including
 optics polishing and support errors, optics misalignments,  
  telescope tracking errors and local dome seeing.  

 The design does produce significant cubic (pincushion) distortion; 
  the image radius/angle relation is well approximated by 
 \begin{equation}
 \label{eq:distort} 
  r(\theta) = 12072\,{\rm mm} \, (\theta + 44 \,\theta^3 - 10\,300 \,\theta^5)
\end{equation}  
  where $\theta$ is the off-axis angle in radians. 
 This implies that 
  objects at the corners of the field are imaged 
  at a radius $\approx 0.8$ percent larger than the linear term;
   thus the (radial) pixel scale $1/ (dr/d\theta)$ 
  varies by 2.4 percent from centre to corner of the field.   
 Pixel solid angles are proportional to 
  $(\sin\theta / r) \, d\theta/dr$, and are thus 3.2 percent smaller at the
   corners of the field.  
  This distortion has no effect
 on image quality, but needs to be accounted for in image processing: 
 there are two effects, firstly a jitter offset of e.g. 10 arcsec
 does not correspond to the same number of pixels at all points 
 in the field.  
 Secondly,  flat-fielding images to produce the standard convention of
  uniform counts per pixel in the sky 
  results in photometric variations proportional
  to pixel solid angles. 
  Both these effects are compensated in the data processing
 software.   

 The science filters are placed close to the detector plane 
  (with a nominal clearance of 15\,mm). This spacing was
 minimised for two
 reasons: firstly, to allow location of 
 the autoguider and wavefront sensor
  units above the filter wheel to minimise electromagnetic
  or thermal interference between CCDs and IR detectors; and secondly 
 to allow the use of ``mosaiced'' filters, one pane of
 glass per detector, 
  without vignetting of detectors by the filter support frame, 
 since monolithic filters of the required size and quality would be 
   extremely challenging or impossible to manufacture. 
 The filters operate in the converging $f/3.25$ beam, with 
  resulting angles of incidence up to 10 degrees. This implies that
 there is a slight passband smearing since interference filters
 have a slight dependence of passband on angle of incidence; 
  the effect is nearly negligible for
 broadband filters, but implies that narrowband filters cannot
  be much narrower than $1.5\%$ bandpass. 

 The design has slight longitudinal chromatic aberration, which is 
 compensated by tuning the filter thicknesses per band
  to compensate: we chose to fix the H filter parfocal to
   10.00\,mm of fused silica; the M2 position was held fixed across
  all passbands, while the filter thickness was allowed to vary  
   for other bands for optimal focus.  The result is that 
   the Y filter is parfocal to 9.65\,mm silica, while
  the Ks filter is 10.37\,mm silica.  We chose {\em not} to allow refocusing
  the secondary mirror between passbands, because the 
  autoguiders and wavefront sensors use fixed filters selecting 
  roughly I-band light, 
  which is folded by a pickoff mirror above the science filter: 
  and refocussing M2 between passbands would have slightly degraded 
   their images and added software complications.   

 Lateral chromatic aberration is small but not negligible, 
  with approximately 1 pixel
  shift between the model J and ${\rm K_s}$   
 image centroids at the corner of the field. 
 The colour-dependent astrometric shift 
  within any single passband is clearly much
  smaller, but this will need to be corrected 
  for in the highest-precision astrometry.   

\begin{figure}[t!]
   \centering
\includegraphics[width=9cm]{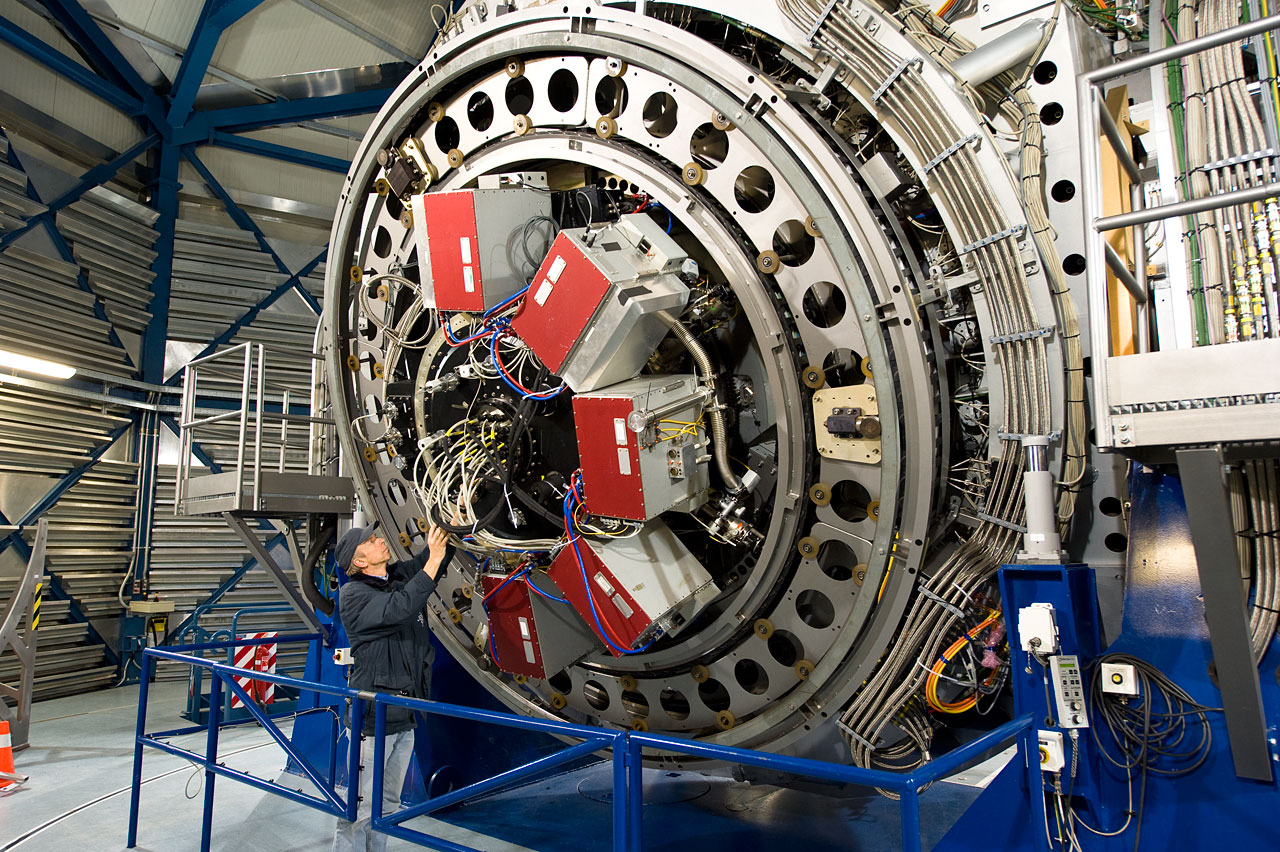}
 \caption{Back-end of VISTA, with the VIRCAM
 at Cassegrain focus. The rear of the cryostat is seen (black), with
 the filter wheel bulge at the far side.  The five red electronics 
  boxes are mounted on the cryostat.  The camera coolers and cryo-pumps
  are mostly hidden behind the electronics boxes.  
 The large annular structure surrounding the camera is the Cassegrain
  cable-wrap; fixed sections of Helium hoses are seen curving around
 the back of the M1 Cell, and enter the cable-wrap at the 
  connector box at lower right.} 
\label{Fig-cass} %
\end{figure}

  The VIRCAM focal plane is located 1172.6\,mm below the pole of
 the primary mirror, while the Cassegrain rotator back face 
  is 725\,mm below the primary pole:
  this gives 447.6\,mm back focal distance  
  from the Cassegrain rotator to the detector plane, which 
  is sufficient to accommodate the 
  large filter wheel bulge and several cryocoolers 
   behind the M1 cell structure  
 (Fig.~\ref{Fig-cass}). 

\subsection{Baffles, stray light and ghosts} 


 Baffling against unwanted light-paths
  from sky to detectors is provided largely by two baffles: 
 the cold baffle tube inside
  the camera cryostat, and the warm ``Narcissus'' baffle around M2. 
 The cold baffle has a front aperture of 812\,mm diameter
 at height 2179\,mm above the detector plane. 
  The warm Narcissus baffle around M2 is a polished aluminium  
   annular baffle of 1.63\,m outer diameter,
  made as two concave nested 
  spherical surfaces. The curvatures of these spheres are
  chosen so that the detectors view mainly
  the black walls of the IR Camera lens barrel 
  reflected in the M2 Baffle. 
 The cold baffle and M2 baffle in conjunction block all direct rays
  from sky or dome to the detectors.   
 The cold-baffle front aperture is slightly ``undersized'' and introduces
 a very small amount of vignetting at the corners of the field
 of view: the vignetting is zero at angles $\le 0.69$ deg, rising to
  1\% at the corner of the field.  This small amount of 
  vignetting was an intentional choice, since the undersizing of the cold
  baffle reduces the required M2 Baffle diameter compared
  to a strict zero-vignetting system, and this reduction produces
  benefits for diffraction and overall average throughput.  
 
Given the absence of an intermediate cold stop, 
 stray light and ghosting were a significant concern 
 (Patterson \& Wells \cite{pw03}), and 
 careful attention was paid to minimising these by design: 
  all surfaces around
 the light beam are shaped to minimise first-order
  scattered light, using various features such
 as ridges on the spider vanes, 
 grooved walls in the camera lens barrel, 
  chamfered edges of the filter mount trays, etc. 
 The VIRCAM corrector lenses are generously oversized with a physical
 radius at least 2~cm larger than their useful aperture; this keeps the
 lens edges ``in shadow'' behind support structure, 
  avoiding stray reflections from 
  the lens outer edges.  
 Additional baffling against moonlight illuminating the
  optics is provided by a movable 
 moonscreen at the top of the slit (Sect.~\ref{sec-encl}). 

 Ghost images from unwanted reflections are present (as
 in any transmissive system); in VIRCAM the dominant ghosts
 are the localised doughnut-shaped out of focus ghosts from 
  unwanted reflection at the filter. The smallest ghost is from
   light reflected twice inside the filter;  brighter stars 
 ($J \simlt 7$) also show
 two larger and fainter ghosts, arising from one reflection
  off the detector surface followed by a second reflection from the 
   bottom or top of the filter. 
 These three ghosts are $\approx 63$, 144 and 205 arcsec 
  in diameter respectively.    
 These filter ghosts were essentially unavoidable, given the several 
 optomechanical constraints requiring the filters to be located close
  to the detector plane; 
 the ghosts are localised around the main star image 
  (though not precisely concentric), and are easy to recognise
 in the images.  To date, very few ``non-local''
  ghosts have been detected\footnote{There is one suspected
  ghost $\sim 3$ degrees from the very bright star Mira Ceti.}, 
  and there is no significant
 ``sky ghost'' or ``pupil concentration'' effect, as can arise
   from reflections between detector and corrector lenses.   

\subsection{No-lens test setup} 

  We note that for testing the telescope without VIRCAM, 
  using the 2 mirrors alone without the VIRCAM field
  corrector (with unchanged M1-M2 spacing) 
 produces unacceptable spherical aberration at the resulting bare focus.
 However, a good ``bare telescope'' 
  configuration with no lenses can be found by allowing
  {\em both} the M2 and focus locations to vary. The solution  
  is to move M2 downward by 2.1\,mm with the hexapod, and locate the 
  detector at the new focus 1153.2\,mm below M1 pole.  
 This setup corrects both defocus and 3rd order spherical aberration,
  and gives a usable field of view $\sim 5$ arcmin diameter, 
   limited by residual off-axis coma since the telescope is
  not an exact Ritchey-Chretien.  
  The 5 arcmin field is ample for test purposes, and was used 
  for the initial telescope commissioning phase
  prior to first light of VIRCAM. 
  In practice, we set up the test camera location by metrology 
   with respect to M1, 
  then moved M2 to best focus, and verified that the spherical
   aberration was minimal as predicted. 

\subsection{Optical design summary} 

 As above, the overall system design is strongly optimised
 for wide-field survey operation with infrequent instrument changes. 
 The design has many beneficial features: 
 the telescope structure is very compact and rigid, leading to a small
  enclosure, reduced costs, fast jitter movements and minimal windshake.  
 The cold-baffle
 camera design enables a very wide field of view with good
 image quality across Z through $\Ks$ bands, excellent throughput
  and no exotic lens materials; and   
 the Cassegrain instrument station can readily accommodate the 
  large and massive VIRCAM and associated service bundles.  

 The telescope also offers good potential for future instrument(s): 
   the large Cassegrain volume implies
 that the potential field of view is limited mainly by manufacturability
  of the new field-corrector, rather than telescope constraints.  
 For the planned 4MOST instrument, 
  corrector designs up to 3 degrees diameter 
  were developed in the concept phase; after 
 cost/benefit tradeoffs the 4MOST baseline is currently 2.5 degrees 
 (de Jong et al \cite{f-most},  see also Sect.~\ref{sec-perf}).   
  
 The main downsides of the design are that the fast $f/1$ primary
  mirror is highly aspheric, and was even more challenging 
 than expected to figure to the required accuracy, 
  leading to unanticipated delays in project completion.   
  Also the relative alignment tolerances of M1/M2 are very stringent,
  which is solved using active optics as 
  detailed in Sect.~\ref{sec-actopt}. Finally, the camera vacuum window is
 very large and gave challenges with manufacturing and thermal design, 
  as in Sect.~\ref{sec-cam}.


\section{Mirrors and Mirror Support} 
\label{sec-mirrs} 

In this section we summarise the physical details of the two mirrors, 
 and their support systems including the M1 active force actuators 
 and the M2 Hexapod. Where needed, we refer to a coordinate system 
 in which $+z$ is parallel to the telescope optical axis, $x$ is 
  parallel to the telescope altitude axis, and $y$ is perpendicular
  to both, i.e. the downward tangent to the primary mirror.

\subsection{Primary Mirror} 

The primary mirror is made of Zerodur glass-ceramic, 
 with the blank manufactured by Schott Glas, Germany; 
 a pre-existing thicker Zerodur blank was machined down by
  Schott Glas to the final meniscus (Doehring et al \cite{schott}). 
 The mirror geometry is a solid meniscus, of  
  $4.10 \,$m outer diameter with a $1.20\,$m 
 diameter central hole, a thickness of 17\,cm, 
  a spherical back surface, and a mass of 5.5 tonnes including the 
 support pads.  
 The measured coefficient of thermal expansion (CTE) is $0.066 $\, ppm/K, 
  small enough that thermal effects on the figure during
 operations are negligible.  
  
 The mirror optical surface was 
  polished by LZOS, Moscow, (Abdulkadyrov et al \cite{lzos-m1}) 
 to the required hyperboloid figure
 as in Table~\ref{tab:mirrs}. 
   This was a challenging and time-consuming process due
 to the high asphericity of the fast mirror (deviation 
  over $800 \mic$ from the best-fit sphere); 
 but was eventually completed to good quality with an rms
  wavefront error of $35\, $nm (after theoretical 
  subtraction of available low-order active force patterns).  
 The figuring process involved numerous cycles of wavefront measurement
 and polishing, roughly once per week for the two year figuring
  process. Wavefront measurements sometimes used two independent null
 correctors of differing design, 
  to guard against the well-known HST-type disaster 
 arising from a manufacturing error in a single null corrector.
  In addition, one of the null correctors was independently validated 
  using a purpose-made computer generated hologram (CGH).  


Mirror coating (silver or aluminium) is described in Sect.~\ref{sec-encl}.   

\begin{figure}[t!]
   \centering
\includegraphics[width=9cm]{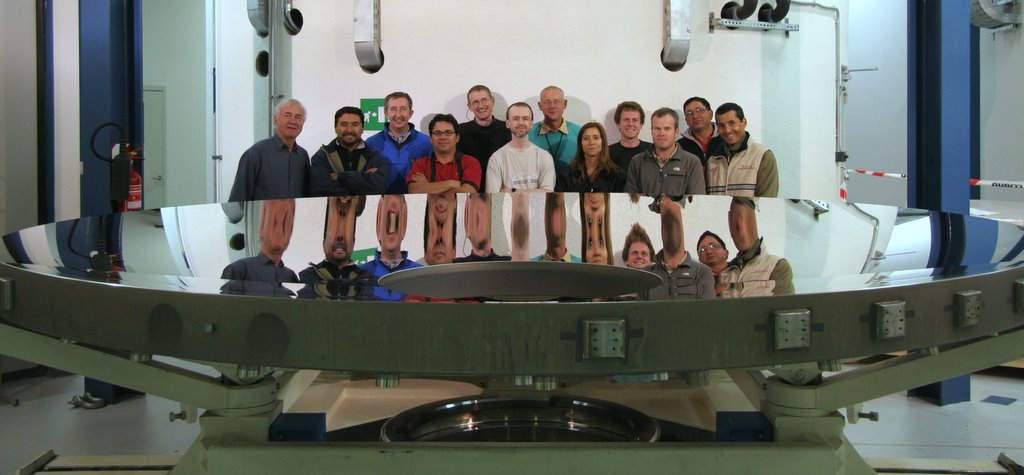}
\caption{VISTA primary mirror on its wash-stand, after the first
  coating in April 2008. The telescope pier is behind the group. The 
  apparent ``kink'' 
  near the front arises from two reflections off the mirror.  
} 
\label{Fig_M1_coated} 
\end{figure}

\subsection{M1 Supports} 

During observations, the M1 is supported against gravity and wind loads 
  by numerous pneumatic {\em force actuators} which deliver
  controlled forces to the mirror back and sides, and it is held in
  position relative to the Cell by six {\em definers}. More 
 details of this system are given by Stobie et al (\cite{stobie}). 

 The definers are quasi-rigid metal cylinders with 
  2-plane flexure joints at both ends, so each definer provides 
  essentially a one-dimensional position constraint:
  thus the six definers combined 
  provide kinematic location of the M1 relative to the Cell, 
  without overconstraint stresses from flexure of the Cell. 
 Each definer's axial stiffness is $30\, {\rm N} \mic^{-1}$, which leads
  to M1 rigid-body rocking frequency above 15\,Hz. 
 Each definer incorporates a load-cell measuring the force component
  along its long axis: 
  these force readings are used as the inputs to a fast servo loop 
 controlling the pneumatic mirror supports, to keep the residual forces
 on the definers at desired values (see later).   

 There are three axial definers (parallel to the $z$-axis) 
  equispaced around the back of M1, and
 three lateral definers attached tangent to the M1 back near the
  outer edge. All the definers
 include force-limiting spring-plate devices, so they ``break away'' and
  become spring-loaded at axial force exceeding $1200$\,N ; this 
 avoids excessive point loads on the mirror and load cells 
  when the pneumatic supports are turned off, 
  or faulty, or in the event of an earthquake. 

  Each of the six definers includes a precision length-adjustment screw
  and dial giving 1\,mm length change per revolution: 
  these are accessible by hand from behind the M1 Cell, 
  enabling the M1 position to be adjusted to $\sim 25 \mic$ resolution. 
  This is essentially a set-and-forget adjustment to align the M1 
   to the Cassegrain rotator axis,  since the Cell structure 
  is designed to deliver minimal flexure between 
  the M1, Cassegrain rotator and Camera (see Sect.~\ref{sec-actopt} for
  more details). 
  Six linear variable displacement transducers (LVDTs) 
  provide monitoring of the M1 position relative to the Cell to
  $\approx 2 \mic$ resolution;  this position information is not used
  in closed-loop, but provides a health-check
  on the support system, and also verifies that any desired 
  manual alignment adjustments have been applied correctly.   
 
 The M1 supports interface to the mirror via super-Invar pads glued to its 
  back and side: 
 there are 84 axial support pads on the back surface, located on four
  circles of 12, 18, 24, 30 supports respectively, giving a six-fold symmetry 
 in the pattern. There are 24 lateral
 support pads around the outer cylindrical face. 
 A further 3 pads on the back face (near the outer edge) provide
  attachment for the lateral definers. 
 For the axial supports, 
  a stainless steel wedge is bolted to each Invar pad
  to provide a flat surface perpendicular to the $z$-axis, against
 which the axial support piston pushes. 
 
%
\begin{figure}[t!]
\centering
\includegraphics[width=9cm]{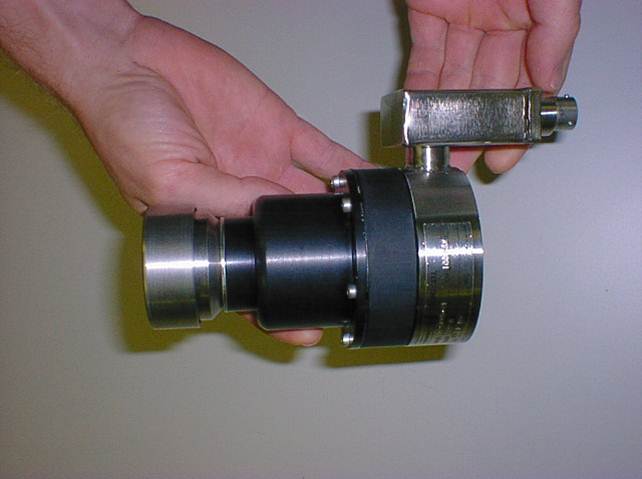}
 \caption{One of the 81 primary mirror axial supports. In operation, 
  the metallic cap (at left) pushes against 
  the M1 metal pad via the pneumatic cylinder at centre. 
 The load-cell is on the right, with readout electronics
  in the cuboid unit above. 
} 
\label{Fig-m1supp} %
\end{figure}

 All the M1 force actuators are pneumatic pistons containing 
  a Bellofram membrane (Bennett \& Baine \cite{bb04}): 
 there are 81 identical axial actuators (Fig.~\ref{Fig-m1supp}), each with
  its own pressure-control valve and load cell, 
 thus all the axial support forces are controllable 
 individually by software.  The axial supports are ``push only'' with an
 operating force range of $5 \newt - 990 \newt$,
  compared to the mean gravity load of $ 645 \newt \, \sin({\rm Alt})$
   per support, and rms force accuracy $\approx 1 \newt$. 
  {\newtwo The air pressure for each support 
  is controlled in closed-loop using the load cell 
  value, and load cell readings are reported periodically to the telescope
  control system for automatic status checking. } 
  Active optics ``pull'' forces
  are simply delivered by pushing less than the mean gravity load,
   so this limits the available active force range 
 to approximately $\pm 250 \newt$ per support. 

   The remaining 3 points in the 84-point axial support
 pattern are the passive axial definers: the fast-balancing servo 
  loop reads the definer
  load forces as input, and adds piston and $x/y$ gradient modes to the
  81 axial supports in order to keep the definer loads at 
   their fair share of the  overall load. 
 This force balancing loop runs fast, at $\approx 20\,$Hz framerate, 
 to ensure that varying wind and gravity loads are rapidly shared out 
  across all the supports, avoiding residuals at the axial definers. 
 The active force pattern 
  adjusting the M1 figure (Sect.~\ref{sec-actopt}) is additive to
 the balancing forces, but this loop runs much more slowly, 
 updating approximately once per minute during telescope tracking.  
 In the event that one axial support is defective, 
  the software can disable it and use
 pre-computed lookup tables to re-distribute its force 
  across the remaining supports with minimal effect on the M1 figure.

 The M1 lateral supports are also pneumatic, 
 and comprise twelve ``push only'' supports below
  the mirror, and twelve ``pull only'' supports above the mirror. 
 The lateral supports are tilted up/down with varying angles
  (roughly tangents to the mirror surface) 
 optimised by finite-element analysis, but all act along lines
 perpendicular to the telescope Altitude axis in the $yz$ plane, 
  i.e. there are no force components parallel
  to the Altitude axis.  
 The lateral supports are not controlled individually, 
  but are fed by four pressure-control valves 
  each feeding one quadrant of six supports: 
  currently all four valves receive one common pressure demand,
  but the use of four parallel valves provides faster response. 
 The lateral force control loop simply servos the pressure demand 
   so that the average of the forces measured by the 
  4 o'clock and 8 o'clock lateral definers is zero;  
  thus the lateral supports balance the component of 
   gravity load (and wind, if any) in the tube $y$-direction. 
  Any small $x-$forces or $z-$moments
  (resulting from deviations from the nominally symmetrical system)  
  are reacted passively by the lateral definers; these residual 
   uncontrolled forces can be checked to be small using the lateral 
  definer load readings.  

\subsection{M1 rest-pads} 

In addition to the active pneumatic support system, the M1 Cell contains 
  a system of 18 passive
  rest-pads which support the M1 when the support system is 
 turned off, or the telescope is parked,  or in an earthquake condition. 
 The rest-pad system comprises twelve axial rest-pads (in two rings of six) 
  behind the M1, and six lateral
 rest-pads equispaced around the side of M1. Also, above the mirror  
  there is a failsafe ``restraint clamp'' which can press down on
  the periphery of the M1 front surface (outside the
 optically useful area). 
 
 The rest pad positions were set as
  a one-time adjustment so that each pad has a nominal 1\,mm air gap
  to the M1 back or side face when M1 is at its operational position. 
 Each rest pad comprises a plastic-foam-steel sandwich, with the plastic
  closest to M1. The 3\,mm thick foam layer provides shock absorbing
  in the event that M1 strikes the rest pads at significant speed,
 which can happen in cases of an earthquake, or a major software fault
  ``dropping'' the mirror. 
  
 The restraint clamp is a rubber-coated steel ring covering
 the outermost 3\,cm of the M1 upper surface (which is unseen by 
  the VIRCAM detectors).  The clamp has a failsafe design: in the power-off
  or parked condition, the clamp is 
  pressed downwards on the M1 by sixteen permanent springs; 
  during observations, the clamp ring is retracted
  upwards above M1 by compressed air pistons which 
 overcome the spring forces. 
 When the telescope Altitude is below 20.5 degrees, or if the
  earthquake sensor trips, the clamp air supply 
  is cut off by hardware, and the spring-loaded 
  clamp presses M1 against the axial rest-pads with 
  $\approx$ 24\,kN downforce.   

 A deliberate omission is that 
  VISTA has no on-telescope M1 cover. This is because
 the tube and camera design do not allow space for a ``concertina'' design 
 (and these are also prone to jamming), while ``petal'' designs 
  were rejected due to degrading the airflow across the primary mirror 
   when open.  
  Various precautions are followed 
  for M1 safety and minimisation of dust buildup:  
  during the daytime the telescope is normally
  parked at Altitude = 20 degrees South, with the M1 clamp applied. 
 A commercial earthquake sensor 
  inside the pier also applies the clamp if an earthquake is detected. 
  A hardware interlock prevents the slit doors moving open unless the
  telescope is in this park position and the moonscreen is deployed 
  (this does not apply to door closing).  
  All lifting operations except M1 removal are 
  performed with the telescope horizon-pointing;  
  an external M1 cover is used during removal of M1 from the
  telescope. 

\subsection{Secondary Mirror} 

The convex secondary mirror (M2) is made of Astrosital ultra-low
 expansion glass-ceramic; optical parameters are in Table~\ref{tab:mirrs}. 
 Both the M2 blank manufacture and the 
  optical polishing (Abdulkadyrov et al \cite{lzos-m2}) 
  were done by LZOS, Moscow, as for
 the M1 polishing.   

 The M2 has a diameter of 1241.5\,mm, central thickness 156\,mm and is 
  approximately 70\% lightweighted, giving a mirror mass of 113\,kg ; this 
  lightweighting was achieved by several months of careful machining 
   starting from a solid block of Astrosital.  
  The M2 face-plate is $\approx 25\,$mm thick, and has an overhang of 
  10\,mm at the outer rim so that the annular baffle 
 surrounding M2 can overlap behind it, avoiding detectors viewing
 sky through a gap.  Since the M2 
  forms the aperture stop for the optical system, 
 the full diameter of M2 is used optically, 
  except for a 0.5\,mm wide bevelled edge around the circumference
  and a flat patch in the centre used as a 
  coarse-alignment target during initial setup at Paranal.  

 The M2 optical testing used a purpose-made Hindle sphere of 2.4\,m 
  diameter, used with the M2 downward-facing supported on the same Cell 
  as used in operations.  

\begin{figure}[t!]
\centering
\includegraphics[width=9cm]{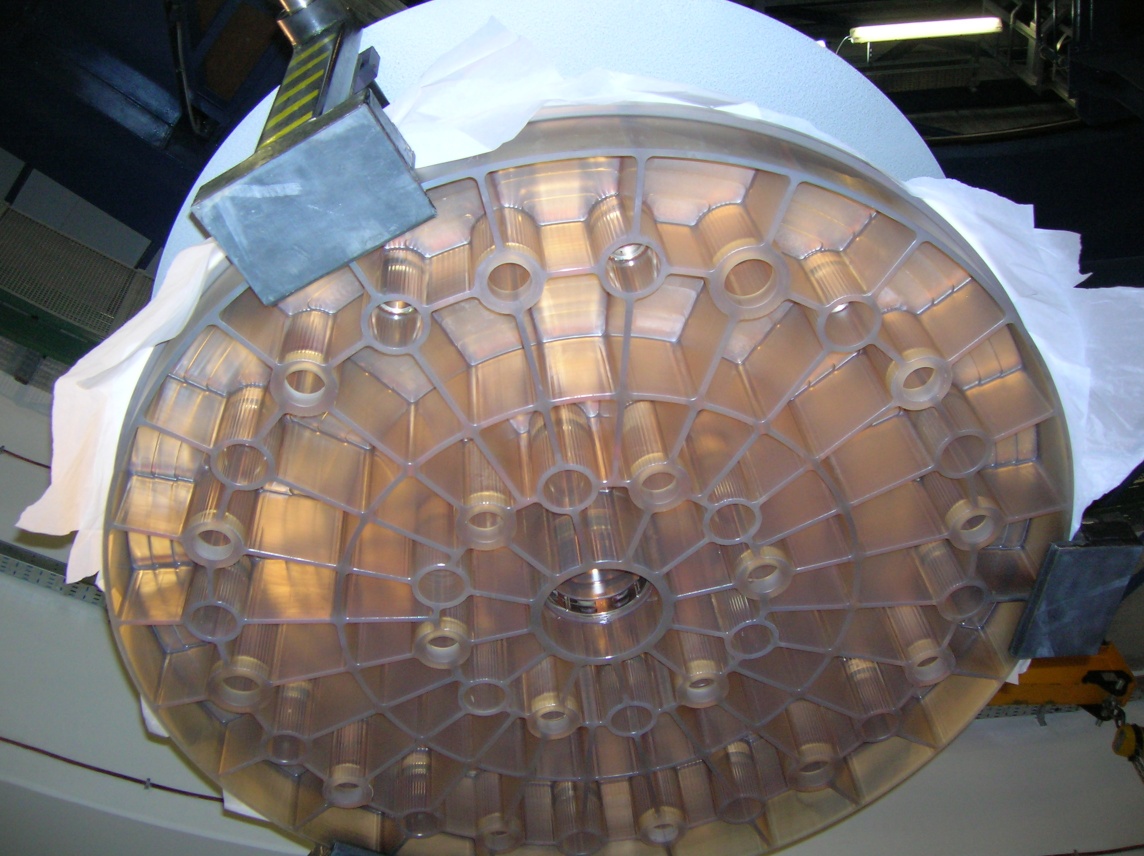}
 \caption{Rear view of the VISTA secondary mirror, showing
 the lightweighting pattern. The astatic lever supports interface to the 
  cylindrical sockets. 
 } 
\label{FigM2back} %
\end{figure}
%

 The M2 is supported from a MgAl alloy Cell 
  containing a passive 36-point astatic lever system.  
 The M2 lightweighting pattern (Fig.~\ref{FigM2back}) includes 
  36 hollow cylindrical pockets machined
  into its back face, in two rings of 12 and 24 respectively; 
   these pockets are alternating between axial and lateral supports. 
  Each pocket has a metal rod attached which passes through a matching
   hole in the M2 Cell. 
  All the 18 lateral supports, and 15 of the axial supports
  have a lever-and-counterweight arrangement, with the pivot
  point attached to the Cell and the counterweight 
  inside the Cell; the lateral supports use a straight rod
  and hinge, while axial supports use an inverted-L arrangement; 
  by basic geometry, these levers deliver lateral and axial forces on 
 the mirror which vary as the desired cos or sin function of altitude 
  (modulo friction).   
  The remaining 3 axial support points are quasi-rigid rods to provide
  rigid positioning of the M2 position relative 
  to the Cell. Lateral centration
 is provided by a large boss on the middle of the Cell 
  with rollers which slide in a matching socket at the centre 
  of the M2 back.

\subsection{Hexapod} 

 In the telescope, the M2 and its Cell are both supported 
  by a large high-precision hexapod:   
  its purpose is to adjust the position of the 230\,kg M2 + Cell 
  assembly in 5 axes (focus, centering and tilt) to a differential
  step accuracy of $\approx 1 \mic$ and $0.1$ arcsec, 
 using information from the wavefront sensors (Sect.~\ref{sec-actopt}),  
  in order to keep the optical system precisely focused
  and collimated under the varying thermal expansion and gravity loads 
  on the telescope tube. 

\begin{figure}[t!]
\centering
\includegraphics[width=9cm]{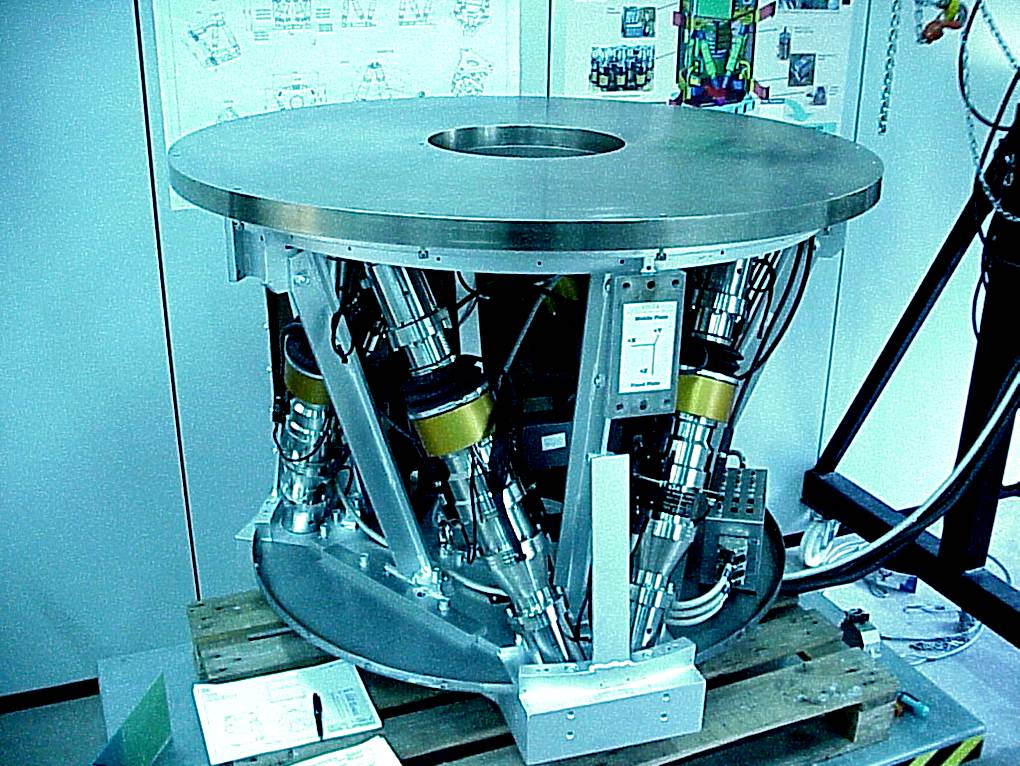}
 \caption{The VISTA M2 Hexapod positioner during testing at NTE, Barcelona. 
 Here the hexapod is inverted from its operating orientation, so the  
  mobile plate is on top. } 
\label{Fig-hexapod} 
\end{figure}

 The hexapod (Fig.~\ref{Fig-hexapod}) was manufactured by NTE, Barcelona 
  (Geijo et al \cite{geijo}), 
  and is a close relation of that used in 
 the Gran Telescopio Canarias 10\,m telescope, with 
 the fast-chopper stage eliminated since VISTA will not observe
 mid-infrared wavelengths $\ge 2.5 \mic$. 
 The hexapod has approximate dimensions of length  90\,cm, 
  outer diameter 110\,cm and mass of 600\,kg.  

 The correction demands to the hexapod are slow, typically
 updating once per minute. Thus, the VISTA system has no high-speed guiding or
  fast tip-tilt correction: windshake rejection relies on
  the combination of the rigid telescope structure and the good 
   wind attenuation provided by the enclosure. 
 
 As the name implies, the hexapod contains 6 legs with
 independent length adjustment: these connect the ``fixed plate'' bolted
 to the Telescope top-end central barrel to the 
 ``mobile plate'' which supports the M2 Cell and M2.  
 Each hexapod leg is made in two parts, coupled by a high-precision
  planetary roller screw of 1mm pitch; a servo motor turns the screw to adjust
 the leg-length. Leg extension is measured in closed-loop by a 
  Heidenhain optical tape linear encoder with resolution $0.1 \mic$, 
  while motor control also uses a rotary encoder. 
 The leg-length travel range is $\pm 5.5 \mm$ to electrical limit 
  switches, and $\pm 8 \mm$ to mechanical endstops. 
 
 The hexapod legs are attached to the fixed and mobile plates via CuBe alloy
  flexure rods: these are sized to allow bending at the joints up to
 10\,mrad
  as the hexapod moves, while providing highly repeatable motion. 
  Safety cylinders around each flexure joint provide an electrical cutout
 which stops motion of all the legs in the event of excessive 
  tilt at any joint. An internal wire rope prevents M2 falling in the event
  of catastrophic failure of flexure rod(s). 

 {\newtwo A fixed framework mounted to the hexapod fixed plate 
  outside the legs supports the outer covers and
 the M2 Baffle, therefore these do not load the hexapod legs
 and the M2 is well shielded from wind loads. } 

 The hexapod is ``slow'' with a response time for small offsets
 approximately 5 sec.  When small position steps are
 commanded by the active optics software, 
  the low-level hexapod control software moves 
  the six legs at proportional rates, 
  so that adjustment steps in focus and centration (rotation around
  M2 centre of curvature) give negligible image shift, and
 these corrections are applied while science exposures are active. 
  In contrast, M2 tilt corrections {\em do} produce image shift, 
  therefore cannot be applied during science exposures:
  these corrections are buffered by the telescope control software 
  so they are only applied in between 
  consecutive science exposures (see Terrett \& Sutherland \cite{ts10}). 

\section{Telescope Structure} 
\label{sec-tel} 

\begin{figure}[t!]
\centering
\includegraphics[width=9cm]{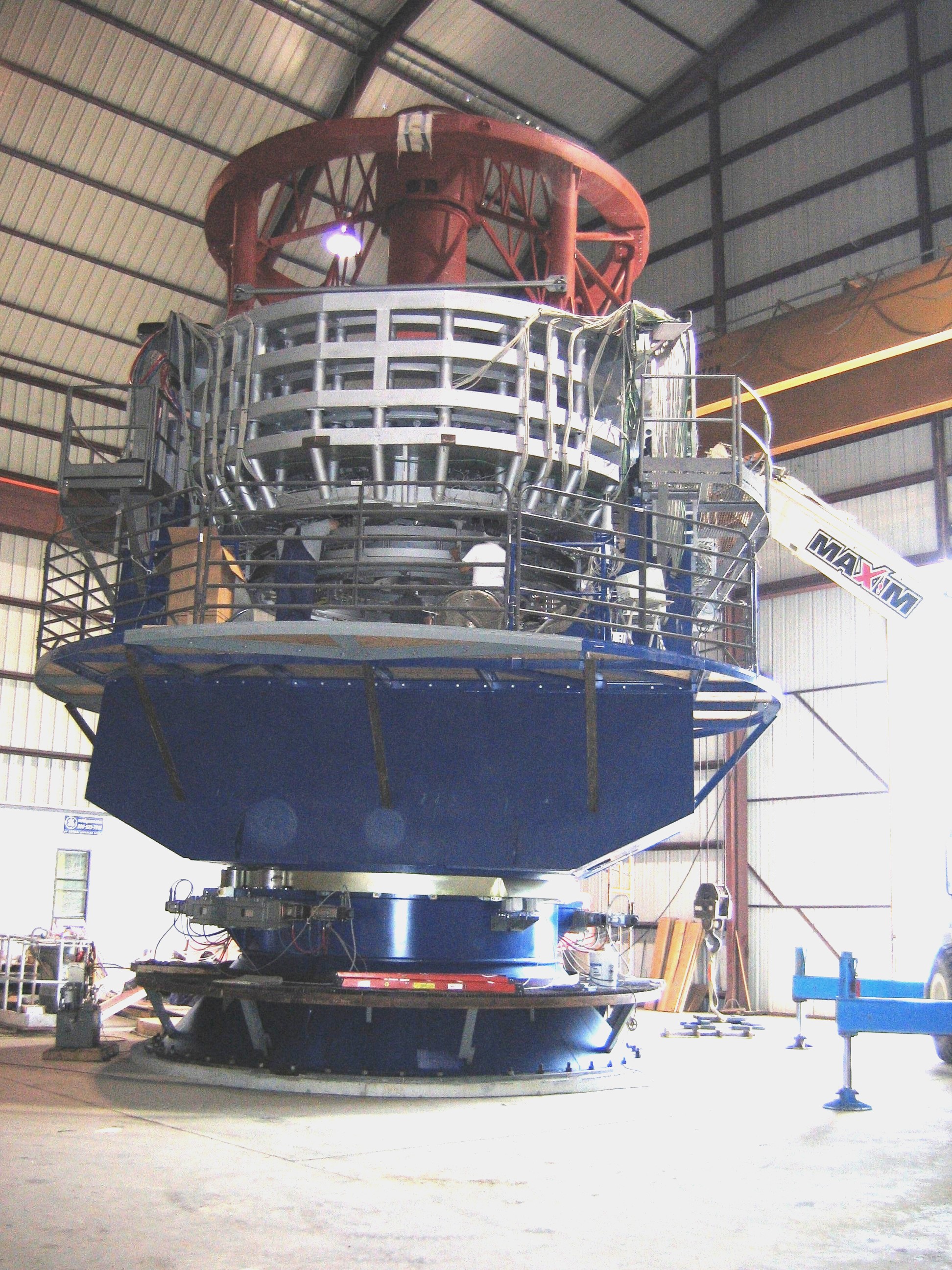}
 \caption{
 VISTA telescope structure in 2005, during 
  testing at the Vertex plant in Mexia, Texas. 
 } 
\label{Fig-tel-fac} %
\end{figure}

 The telescope structure and axis controls 
 (and also the M1 support system) were
 manufactured by Vertex-RSI (now General Dynamics) of Mexia, Texas; 
  many details are given in Jeffers et al (\cite{jeffers}).  
  Much of the altitude and azimuth bearings and drive systems re-use the 
  design of the 4\,m SOAR telescope at Cerro Pachon, 
 though the optics and tube structure of VISTA are very different.  
 A view of the complete telescope structure during factory testing is
  shown in Fig.~\ref{Fig-tel-fac}. 

The key design aims of the telescope structure were
 to provide good tracking performance, 
 to maximise stiffness (for windshake rejection), to
  minimise deflection between the M1 and Camera during operation, 
 to provide fast offsetting for jitter movements during observing, 
  and to provide fault-tolerance and easy maintenance for
 low operation costs; these have generally been successful. 
 
 One notable feature is that the telescope uses rolling-element
 bearings on all three rotation axes: these have slightly higher
 friction than the hydrostatic bearings on most modern large
 telescopes, but offer performance that meets 
 our specifications at lower cost and significantly reduced 
 maintenance, since high-pressure oil pumps are not required. 

 The telescope tube does not use the traditional Serrurier truss principle:
  instead, to maximise overall stiffness 
  the M1 Cell is rigidly coupled to the central Altitude ring, 
 and the gravitational deflection of the top-end 
 with varying altitude ($\sim 0.4\,$mm max) is compensated in closed-loop 
 by the active optics system and the M2 hexapod. 

The overall Telescope moving mass 
 is 90 tonnes, including mirrors and VIRCAM:  this divides as 
  44 tonnes in the ``tube assembly'' rotating around the altitude axis,
   and 46 tonnes in the azimuth structure. The lowest natural
 frequency of the structure including the pier is 9.2 Hz, 
 showing good overall stiffness. 
 
The telescope structure comprises various subsystems, as follows: 

\subsection{Azimuth rotation system} 
 The telescope is supported on a cylindrical concrete pier,
 of height 3.66\,m, outer diameter 6.0\,m and inner
 diameter 4.0\,m. 
  The hollow centre of the pier contains 
 the hanging-spiral azimuth cablewrap, accessible 
 via a door at ground level and internal ladders.   
 {\newtwo A central pillar of 1\,m diameter supports the azimuth encoder
   tape disk above. }   
  On top of the pier is the fixed pedestal,
   a 20 tonne steel cone structure: 
  this pedestal provides a precision flat upper surface supporting
  the telescope azimuth rotation bearing, and rigid mounts 
 for the azimuth drive gearboxes. 

 The azimuth structure rotates on a rolling 
   element bearing of 3.66\,m main race diameter.  
 The azimuth rotation is driven by a set of 
  four identical motor and gearbox systems, operated 
  in two counter-torqued pairs to minimise backlash. 
  Each motor (a commercial Kollmorgen DC servomotor) 
   drives a commercial Bayside 70:1 planetary 
  high-speed gearbox, then a custom-made 10.65:1 low-speed 
  right-angle gearbox fixed to the pedestal.  
  Each low-speed gearbox turns a 37-tooth helical pinion which engages with the
   728 teeth machined on the outer face of the azimuth 
  bearing ring; thus, the total gear ratio is 14\,671:1, or 88.3 arcsec 
 telescope rotation per motor revolution. 
  This large gear ratio provides fast acceleration and offsetting performance; 
  the maximum slewing speed is 2 deg/sec, limited by the 5000\,rpm 
   limit of the motors.  

 The Azimuth bearing supports the yoke, 
  a large U-shaped steel plate structure, 
  of mass $\approx 32$ tonnes, which rotates in azimuth and 
  supports the telescope altitude bearings on its upper faces. Hatches
  are provided in the yoke for access to the various cables and hoses.
 The yoke also supports the circular rotating floor 
 forming the central part of the Dome floor, and two side platforms 
 giving easy access to the Altitude drives.  

\subsection{Altitude rotation system} 

The Altitude rotation system is fully symmetrical, with identical
 bearings, drive motors and encoders on both sides of the 
 telescope tube, to minimise structural torquing effects. 
 The two altitude bearings are large rolling-element bearings; 
 each side uses a pair of preloaded bearings using conical rollers, 
  outer diameter 920\,mm. 
 The altitude axles are hollow, enabling safe access 
  to the M1 side supports by
  removing the motor cover and crawling through the 630\,mm bore.  
 The altitude axis is driven by a pair of identical direct-drive motors 
  adjacent to the bearings.  

 The altitude axis motion has several layers of safety features to avoid
  out-of-range collisions: 
  firstly, velocity limit switches constrain the maximum 
  velocity to 10\% of slewing speed
  as the axis position approaches a hardware limit. 
 Next, pre-interlock switches command the motor drives to
 back away from the limit under hardware control. Next, 
  interlock switches cut the motor power and apply the failsafe
  air-brakes. Final protection is provided by two pairs of 
  large oil-filled dampers mounted on the yoke; if the tube
  travels out of range to Altitude 
  $\approx 92^\circ$ or $-3^\circ$, the back of the M1 Cell hits
  one pair of dampers, which are calculated in theory
   to stop the tube from 4 deg/s (twice the operational limit) 
  without causing damage.  
  For prevention of overspeed in the event of control failure, 
  a small gyroscope fixed on the tube interlocks the drives 
  if the tube angular speed exceeds 2.5 deg/s. 

\subsection{Telescope tube} 

The telescope tube assembly, supporting the optics and rotating
 in altitude, comprises
 three main parts: the M1 Cell, Altitude ring and Top-end structure. 
 The 15-tonne Altitude ring and 12-tonne M1 Cell were manufactured in separate 
 weldments for ease of transport, but were permanently bolted together
  during installation on site:   
 these form a rigid ``bucket'' structure which connects the 
  M1 definers and the Cassegrain rotator to the Altitude rotation 
  axles.  Both parts are a lattice structure of steel tubing,
 to provide improved ventilation of the primary mirror
  and access to the M1 support system. 
 Since the Cassegrain rotator axis forms the fundamental alignment
  reference of the system, it is helpful that this is never
   detached from the body of the telescope. 


  For optimal alignment, 
  the telescope structure had specific and stringent specifications
 for the relative deflections between
 the M1 and the Cassegrain rotator. In order to meet these, 
  a novel structure for the Cell was used: 
  the lower section contains a rigid triangle beam
  which links the M1 definers to the Cassegrain rotator,
  while a separate relatively lightweight 
  ``basket'' in the upper part of the cell carries the M1 axial supports. 
  This means that flexure of the basket is absorbed by
  length changes in the pneumatic 
   M1 supports (which run in a fast force-control loop), 
  while the passive M1 Definers keep 
  the mirror position quasi-static relative
  to the triangle beam;  this minimises relative deflections
  between the Camera and the M1, for optimal image quality.  
 Details of the flexure measurements are provided in Jeffers et al 
  (\cite{jeffers}). 

The telescope top-end structure is a 6.5 tonne assembly
 comprising trusses, top-ring, spiders and the central
 ``barrel'' structure. The complete top-end structure
  (without M2) can be detached from the Altitude ring 
  and removed in one piece (horizon-pointing), then stowed on a 
 cradle on the dome floor for M1 removal and 
 recoating (see Sect.~\ref{sec-encl} for details). 
 
 The top-ring has an inner diameter of 4.5\,m 
  (which is sized to not vignette a 4 degree diameter 
  field of view, in case of future instrument enhancements, such as a 
  possible wide-field spectrograph).  To maximise
 stiffness in the focus direction, the spiders 
 are slanted upwards, and are non-radial, joining
  the central barrel at two opposite points.  
 The spiders (each 36\,mm total width) 
 incorporate several small ridges on their side faces, 
 to minimise grazing-incidence stray light 
  scattering off the spider sides and reaching the
  detectors.   

 The top-end central barrel is a passive steel structure, 
  providing a rigid link between the spiders and the Hexapod 
  fixed plate: it has a diameter of $\approx$ 1.4\,m 
  and mass 1 tonne.  The barrel is hidden from the detectors
  by the 1.63\,m M2 Baffle below. The barrel
  has a central hole of 25\,cm diameter, used for early on-sky testing 
  with a 20\,cm telescope mounted on the Cassegrain axis.  
  A recess on top
  of the barrel houses the Hexapod electronics control box; 
 this box is glycol-cooled, via feed hoses running in an insulated
  channel along the top of one spider leg. 

\subsection{Cassegrain rotator} 

As for any altazimuth telescope, it is essential to compensate
  for image rotation around the optical axis as the telescope
  tracks the sky.  
The Cassegrain rotator is attached in a recess on the back of the M1 Cell,
 and forms the main telescope to camera interface: 
  this rotates the 2.9 tonne VIRCAM around its axis.  
 The rotator is driven by one pair of counter-torqued motor and
 gearbox systems, which are recessed into the M1 Cell {\newtwo and drive
  a helical gear on the outer face of the Cassegrain rotator}.  
 The rotator bearing is a ball-bearing with 1.45\,m race diameter;
  the central bore is 1.35\,m diameter, to allow the 
 primary mirror lifting tool to pass through during M1 handling.  

\subsection{Telescope drives and encoders} 

 As noted above, the telescope axes are driven via counter-torqued 
  gearboxes for Azimuth (four) and Cassegrain (two), while
  two direct-drive motors drive the Altitude axis. 
  Angular position measurement is provided by Heidenhain ERA~880C optical
  tape encoders on all three axes: 
 the Azimuth and Cassegrain axes each use a single full-circular encoder
  tape with four equispaced read-heads, while   
  the Altitude axis has a pair of semicircular optical tapes
 on either side of the tube, with two read-heads (fixed to 
 the yoke) at $\pm$ 45 degrees from the vertical on each side. 

 The encoder tape diameters are respectively 1638\,mm, 1854\,mm and 1578\,mm 
  for the azimuth, altitude and Cassegrain axes.  
 The encoder tapes have grooves with 25 lines/mm; each readhead
  provides ``sin'' and ``cos'' signals which are 
   electronically interpolated 
  to 12 bits (4096 counts) per groove, and averaged
  in the electronics;  thus one encoder 
 least-significant-bit corresponds to approximately 10 nm at 
 the tape,  or $\sim 0.0025$\,arcsec around the axis.  

 The control system for each axis uses an outer position loop 
  running in the axis LCU (Sect.~\ref{sec-soft}) 
  based on comparing actual and demanded encoder readings;
 this outputs an analogue velocity demand to the low-level axis 
  control electronics, and in turn this 
  sends current demands to the power amplifiers 
  driving the motors.  
 Since these inner loops are fast and the axis encoders have very high
  resolution,  unwanted effects such as motor torque ripple, 
   gearbox imperfections and backlash are
  largely eliminated by the servo loops, and the telescope
 tracking performance is good  
  (see Sect.~\ref{sec-telsum} for tracking errors).

\subsection{Cable wraps} 

The VISTA cable-wraps are substantial, due mainly 
 to the 10 high-pressure Helium hoses feeding the IR Camera, in 
 addition to many signal cables and glycol coolant hoses. 

 The Azimuth cable-wrap is a hanging spiral inside the pier, 
  using metal hoops
  suspended on chains to offload the self-weight of the cables 
 and prevent tangling: 
  this wrap is simply pulled around by the telescope drives. 
  The Altitude cable wraps (one
  on each side, with separate sides for power drives and signals)
  use Igus energy-chain devices, also pulled
  by the main telescope drives. 

 The Cassegrain cable-wrap assembly is a large annulus
  attached to the back of the M1 Cell; of mass 
  1000 kg, outer diameter 4\,m, and inner diameter 2.7\,m to
  give clearance for the VIRCAM filter wheel and electronics boxes. 
 This cable-wrap is a horseshoe-type design, allowing 1.5 revolutions
 of the Camera end-to-end, therefore allowing any user-selected
 Camera position angle on the sky followed by a 180 degree 
 observing track. 
  The large size of this wrap requires its own independent drive system 
  to eliminate drag torque on the Instrument rotator axis.  
 The Igus cable chain is supported between two frames: 
 an outer frame is fixed to the M1 Cell,
  and an inner frame is mounted on a rotary bearing (separate
  from the instrument) driven
  by its own motor. The cable-chain is held captive between the two frames
  by a wheeled guide drum, which is dragged around by the U-bend in 
   the cable chain, at approximately half the rate of the inner ring.   
  The cable-wrap inner frame is electronically 
  slaved to follow the Instrument rotator 
  angle by an LVDT sensor mounted between the two; 
   lanyard pull-switches stop both systems if the cable-wrap to Instrument 
  following error exceeds $\approx 3$ degrees.  
  Connector boxes are provided on both ends of the wrap, so if necessary
 the complete Cassegrain cable-wrap can be removed as a single
  unit (after attaching bracing bars). Normal removal of the VIRCAM
  leaves the cable-wrap in place.  

\subsection{Telescope summary} 
\label{sec-telsum}  

 The overall telescope tracking and pointing performance is very good: 
 the typical observed high-frequency tracking errors (measured
  from the difference between actual and demanded encoder readings) 
  are $\approx 0.09$ arcsec
  rms around the Altitude and Azimuth axes, and $0.3$ arcsec around 
  Cassegrain axis; 
  note that for conversion from axis to on-sky errors, the Azimuth error
  is reduced by a factor of $\cos({\rm Alt})$, 
  and the Cassegrain error is reduced by a factor of $0.03$ (the field
 diameter in radians); thus for typical airmasses 
 the Altitude errors contribute slightly more than Azimuth, while Cassegrain
  errors contribute much less.   
 Low-frequency errors (e.g. pointing model residuals, thermal drifts) 
  are largely eliminated by the autoguiders.  
 
 Absolute pointing errors have been measured at $\approx 1.0$ arcsec rms
  in full-sky pointing runs at Altitude $\ge 25^\circ$.  
  In practical operation, small non-repeatabilities
 in the M1 position from night to night can degrade this slightly,
 but the pointing is generally good. {\newtwo (The telescope
  normally operates with ``closed-loop'' pointing
  at each target acquisition, by applying a correction 
  to move the guide star onto the theoretically predicted pixel
 on the guide chip). }   
 Differential open-loop pointing accuracy for sub-degree sized offsets
   is comfortably better than 1 arcsec. 

The telescope offsetting performance is excellent, with slew-and-settle
 times of $\approx 7$\,s for offsets $\sim 10$ arcmin, and about 10\,s 
 for degree-sized moves; the structure settles
  fast, with no sign of image oscillations 
  after an offset movement.  

For large-angle moves, the overheads are dominated by maximum 
  slewing speeds, which are $2 \rm \, deg \, s^{-1}$ for azimuth and altitude,
 and $3.6 \, \rm deg \, s^{-1}$ for Cassegrain; axis 
 accelerations are $1 {\, \rm deg \, s^{-2}} $ for azimuth and altitude,
 and $1.8 {\, \rm deg \, s^{-2}} $ for Cassegrain, 
  thus each axis can accelerate 
  from standstill to maximum speed in only 2\,s. 
 These values are fast enough that 
  the time overhead for a large-angle slew is
   usually dominated by the maximum speed of the azimuth axis.  

No significant windshake has been observed; the lowest natural
 frequency of the structure was modelled to be 9.2 Hz, 
  which is among the best for any telescope of similar 
 aperture.  

In summary, apart from occasional transient electrical glitches 
 (e.g. defective power supply units or poor connections) 
 affecting the control systems, 
  the performance of the telescope structure and axes is very good.   
 The M1 support
 system has also been very reliable, after fixing some 
 infant-mortality problems in the force control electronics.

\section{IR Camera} 
\label{sec-cam} 

\begin{figure}[t!]
\centering
\includegraphics[width=9cm]{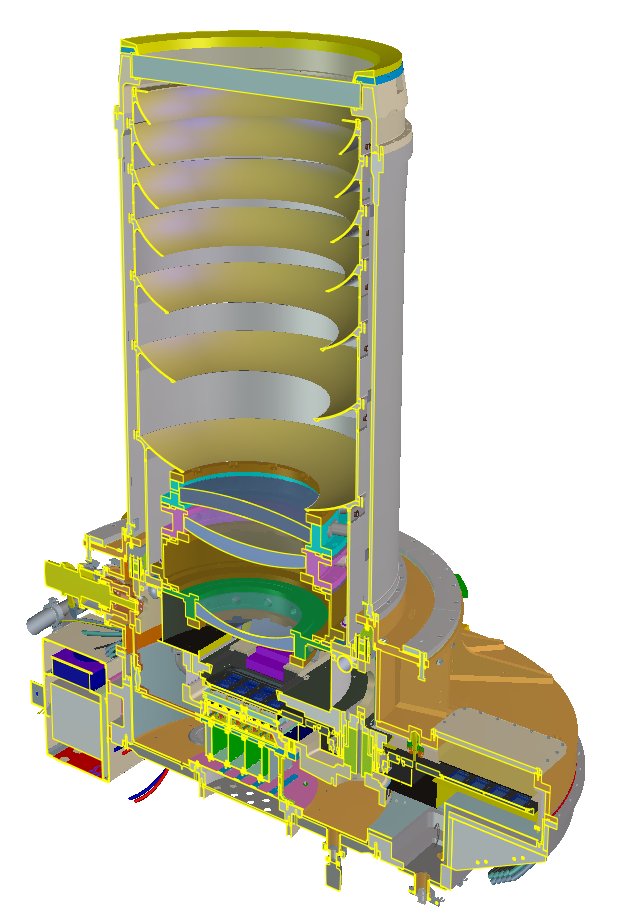}
 \caption{
 A cut-away computer model of the VIRCAM (with some thermal shields 
  and other details omitted); here the vacuum window is at
 the top and the filter wheel is at the bottom right.   
 The ellipsoidal cold-baffles are seen in the upper half; 
 lenses and window are shown in grey. A wavefront
 sensor unit is in purple, and some filters are seen as dark blue 
 squares, with the detectors just below. 
 } 
\label{Fig-cam-cad} %
\end{figure}

\begin{figure}[t!]
\centering
\includegraphics[width=9cm]{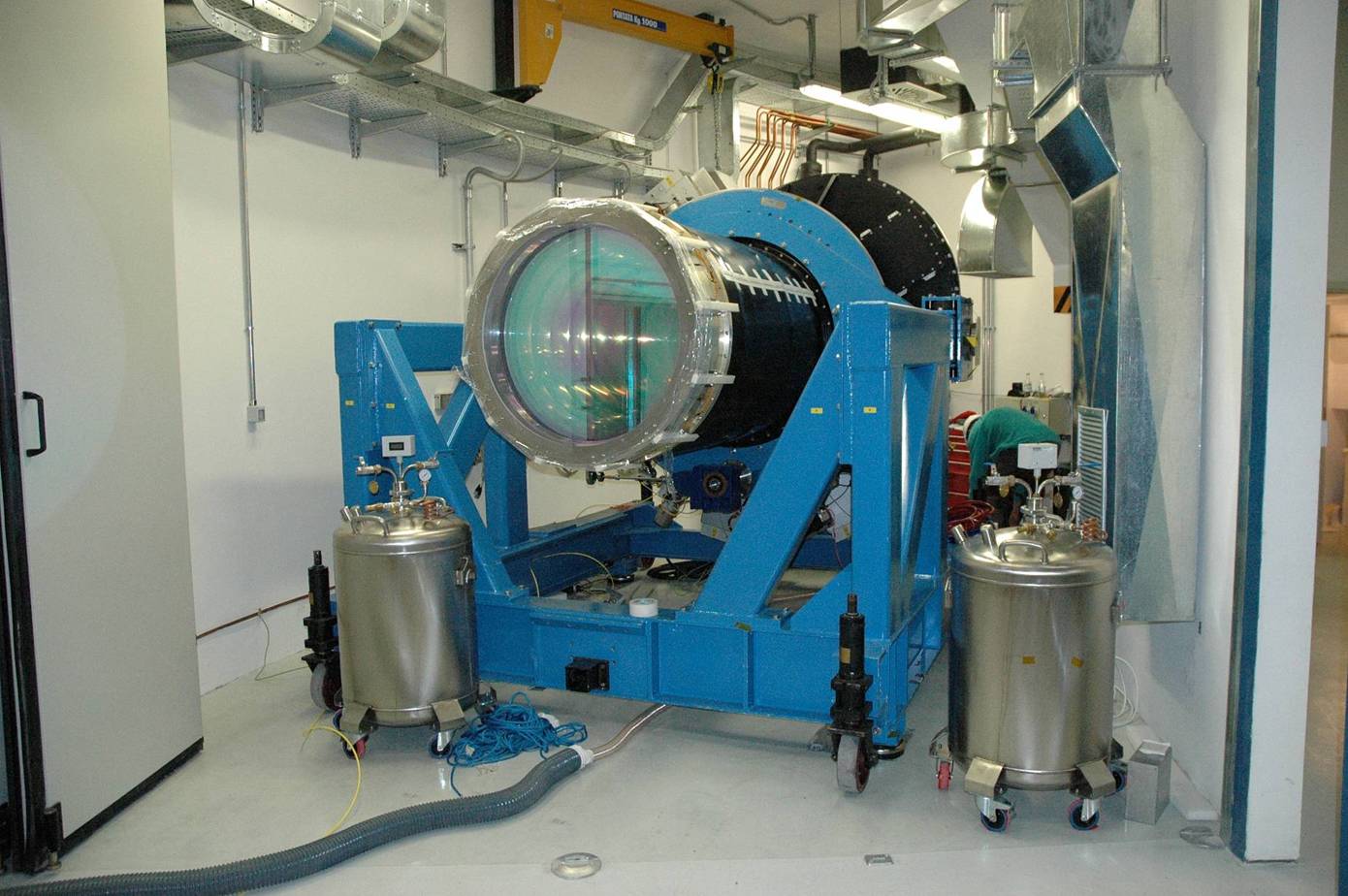}
 \caption{VIRCAM instrument at Paranal, 
   in its lab on the enclosure ground floor.  
 The camera (black) is mounted on its blue handling trolley; the filter
  wheel bulge is visible at the back.  
  Here a transparent protective cover is mounted over the window. 
 } 
\label{Fig-cam-lab} %
\end{figure}
 
The VISTA Infrared Camera (hereafter VIRCAM; Figs.~\ref{Fig-cam-cad},
 \ref{Fig-cam-lab})
  is currently the world's
 largest and widest-field astronomical near-IR imager,
  with a mass of 2900 kg, cryostat length of 2.8\,m, and a corrected field of
  view of $1.65$ degrees diameter. The focal plane contains 
  16 Raytheon VIRGO HgCdTe $2048^2$ detectors, totalling 67 Mpixels, 
  giving an active field of view of $0.60 \,{\rm deg}^2$. 
 Further details of the design and assembly are given by Dalton et al 
 (\cite{dalton}), and commissioning results in 
 Dalton et al (\cite{dalton10}). 

 The cryostat geometry is approximately a stepped cylinder, with
 diameter 1.2\,m at the back end, and 1.1\,m at the front; 
  a large bulge in the cryostat near the back encloses the filter wheel. 
 The camera is mounted to the Cassegrain rotator on the back of
  the primary mirror cell, via a flange at approximately
   80\,cm above the cryostat back-plate.  

 After reflection from M1 and M2, the converging
  $f/3.25$ beam enters the front of the camera via 
  a 95 cm diameter Infrasil vacuum window (located approx 1\,m
 above the primary mirror).  The beam then passes through three
 Infrasil field-corrector lenses, then the filter, 
  and finally reaches the science detectors. 

 Although the camera is unusually large, it is mechanically
  quite simple, with only one internal moving part (the filter wheel).
 This was a deliberate design decision, since the combination of
  the single--instrument telescope 
 and the long Camera warmup/cooldown time implies that any fault 
  requiring opening up the cryostat will 
 lose at least 8 nights of observing time. 
  (Notably, the autoguiders and wavefront sensors were designed with
  no moving parts).   

\subsection{Thermal and vacuum design} 

 The camera internals, totalling
  $\approx 800\,$kg cold mass, 
  are held at operating temperature 
  by three Leybold 5/100T {\newtwo two-stage} closed-cycle coolers; these are 
  fed with Helium gas at ambient temperature and 20 bars by compressors 
 located on the
  ground floor, via hoses running through all the telescope cable-wraps. 
 The vacuum is maintained by two Leybold cryo-pumps (one 
  operating and one redundant), using another two identical compressors: 
  each compressor has a separate supply and return hose, 
  thus there are ten Helium hoses in total running to/from the camera. 
 The cryo-pumps have gate-valves so (if necessary) the getters 
 can be warmed up and outgassed with an external pump 
 without breaking vacuum on the main cryostat. 
 {\newtwo The Helium circuits can be 
  diverted to the Instrument lab via T-valves.} 

 The thermal design of the camera (details in Edeson et al \cite{edeson})
  is essentially a 4-layer Russian doll, as follows: 
\begin{enumerate} 
\item 
  The outermost layer is the vacuum vessel and window.  
  The cryostat nose (above the M1) and window edge 
   are held at ambient temperature by low-power heater
   elements on the vessel, to minimise adverse local-seeing effects. 
\item 
  The second layer is a set of overlapping stainless steel radiation shields 
   attached
  to the inside of the cryostat by insulating standoffs; 
  these shields cool passively to $\approx 240$\,K, and reduce 
   parasitic heat radiating to the cold
  structure. (Use of multilayer insulation was considered but rejected,
  due to concerns about outgassing from trapped volatiles). 
\item 
 The third layer is the main cold structure
  operating at $\sim 100 $\,K : this includes several
  linked Aluminium alloy support frames, to which the 
 lens barrel, cold baffle, filter wheel and detectors
 are all attached.  The central ``optics bench'' 
  is coupled to the cooler first-stage cold heads via 
   large copper braids.
 Most of the cold structure operates at 100 - 110 K, though there
  is a gradient up the cold baffle. 
  The cold structure includes a cover which encloses
   all of the filter wheel;  
  since the corrector lenses are opaque at $\lambda > 3.5\mic$, 
  the last lens cools to $\sim 110$\,K ; thus the filter wheel
  sees only cold surfaces, and cools radiatively
  to approximately the same temperature as the cover. 
\item  
 The fourth and innermost thermal layer is
  the detector system, including the detector mounting plate
 and a copper ``thermal plate''.  This thermal
 plate is linked by copper flexi-straps to the cooler second-stage 
 cold heads, and the detector cooling straps are connected to the 
  thermal plate.  
 Without heating, the thermal plate would overcool to $\sim 40$\,K;
  a servo heater on the thermal plate controls its temperature
 to a setpoint of typically 68 K via a fast inner control
 loop on a Lakeshore controller. This thermal-plate setpoint
 is adjusted in software 
 using a slow outer control loop to maintain the mean detector
 temperature at a user-defined setpoint, normally 72\,K. 

 Detector temperatures show quasi-static differences  
  $\sim 1$\,K between detectors; the more important
  time variation is dominated by a $\sim +0.1$\,K temperature hump 
  for a few minutes after a large rotation of the filter wheel; 
   this occurs because out-of-beam filters sit in the
  cryostat bulge and are modelled to equilibrate $\sim 10$\,K warmer than 
 the in-beam filter, so the detectors experience a slight step
  in radiative heating after a wheel move. 
 The thermal control loop compensates this, but with some lag.  
\end{enumerate} 

 The cold structure is supported from the cryostat vessel
 via a ``crown'' structure of eight G10 glass-epoxy bipods; 
 these provide good strength and low thermal conductivity. 
 (Of the total parasitic heat into the Camera, 
  in practice over 95\% of the heat is
  radiated, and $< 5\%$ is conducted down the bipods.) 
 Each bipod leg comprises 3 blades which are thin in the radial
 direction but very rigid in the tangential
 and axial directions: this allows for thermal contraction of the
  Aluminium cold-structure, while maintaining rigid 
  relative positioning between the cold mass and the cryostat.   

 For cooldown from room temperature, 
  it is possible to cool the camera using the
 closed-cycle coolers alone, but this takes approximately 7 days 
  which is undesirably slow.  
 To reduce cooldown time, a pre-cool pipe is permanently
  attached to the optical bench via conductive links; for pre-cooling,  
  liquid nitrogen (LN2) is fed into this pipe using
 a semi-automated external control device, and boiled off to outdoors;
 this reduces the total cooldown time to around 2.5 days 
  using approximately 400 litres of LN2. 

The cold structure includes a 500 W warm-up heater 
 attached to the Optics bench 
 (fed by separate external power)  which 
 can warm the cold mass to room temperature in approximately 2 days. 

 Also, a dedicated battery supply is provided which
  feeds only the focal plane heater; 
 in the event of a long-duration power loss which 
  runs out the main UPS batteries, or loss of computer thermal control, 
 this backup battery trickles approx 10 W of heat to the detectors 
  thermal plate for up to 48 hours; 
 this ensures that the detectors do not become 
  the coldest point in the cryostat and become a cold-trap
 for outgassing contaminants. 

\subsection{Vacuum Window}

 The VIRCAM vacuum window was probably the third most challenging 
  component in the project, after the M1 and the 16 near-IR detectors.  
 The window is nominally flat, 
  made of Heraeus Infrasil IR-grade fused silica, 
  95\,cm diameter and 79\,mm thick,   
 to provide a very conservative safety margin against the 
 large atmospheric pressure load (60\,kN during testing in the UK, 
  then 45\,kN at Paranal).  

 The window size exceeded the maximum diameter available from Heraeus, so
 a smaller but thicker ingot was provided by Heraeus; this was
 then shipped to Corning, USA for ``flowing out'' into a larger thinner 
 blank, followed by machining to final size. 
 The blank then went to Sagem, France for polishing. 

Several mishaps were encountered: the first two ingots cracked during
 cooling, but the third was successful. After the flowout procedure,
 a handling accident broke a chunk from the edge of the blank; 
 fortunately this chunk was outside the final diameter,
  so this disappeared in machining to final size.   

The refractive index inhomogeneities (probably due to the flow-out 
 procedure) are non-negligible given the thick
 window: this was compensated in the final stage of polishing 
 using a measured refractive index map. 
  Sagem performed a final correction stage by 
 first polishing the surfaces optically flat, then 
 ion-beam polishing non-flatness onto one surface with the opposite sign
 to the refractive index variations.  

The window is broadband anti-reflection coated, and is supported
 on a 91\,cm diameter O-ring in a purpose-made Cell. 
Small heater elements are attached around the edge of the window,
 which are servo-controlled to keep the edge close to ambient 
  temperature. 
 
\subsection{Lens Barrel} 

The VIRCAM contains a 3-lens all-Infrasil field corrector; {\newtwo blanks
 were provided by Heraeus, and optically polished by Sagem, France}. 
  The lens barrel includes alignment shims, so that the final lens spacings 
 were re-optimised after polishing of the lenses 
 based on the measured radii of curvature (Leclerc et al \cite{leclerc}). 

\begin{figure}[t!]
\centering
\includegraphics[width=9cm]{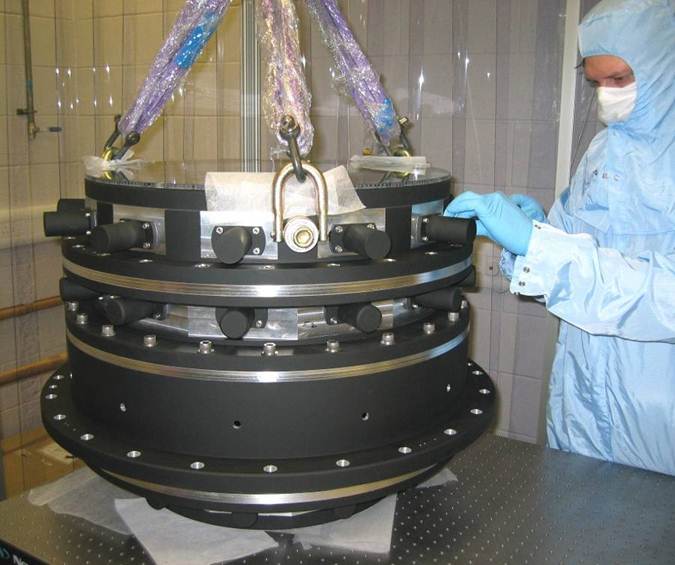}
 \caption{
 VIRCAM corrector lens barrel during final assembly at UKATC,
  with a protective cover on top.  The flange near the bottom 
  mounts to the VIRCAM main optics bench.  
 The cylinders extending radially out from each cell enclose the 
   PTFE thermal compensation rods and springs. 
 } 
\label{Fig-barrel} %
\end{figure}

 The lenses are large, with the largest lens 582\,mm outer diameter,
 therefore their cryogenic mounting required significant design attention. 
 The lenses are supported in an Aluminium alloy barrel
  (Fig.~\ref{Fig-barrel}) to match the remainder of the Camera 
 cold structure. 

From room temperature
 to 100 K, the Aluminium barrel contracts by approximately
  $0.34\%$ while the Infrasil lenses contract by $< 0.01\%$, so 
  thermally compensated mounts are essential to avoid large
  pinch forces on the lenses when cold.   
 Each lens is held 
  centered in its cell by 12 radial rod+spring retainers equispaced
 around its circumference: the rods are made of PTFE  
 which contracts by $1.75\%$ over the above temperature range, 
  relatively about 5 times the barrel. Thus, setting the
  rod lengths to 1/5th of the lens radius
  means that the linear contraction of lens+rod 
  cancels that of the barrel, so the
  spring compression is almost constant with
  temperature.   The springs are preloaded warm to 
  provide positive centration of the lenses, 
   and are stiff enough that the lenses decentre 
  $\le 70 \mic$ under their own weight.  

 We note that using rods alone
 did not ensure positive centration of the lenses, 
  while using springs alone was not acceptable: 
  if the springs are stiff enough  
 to keep gravity decentre $\le 70 \mic$, an uncompensated
 system would induce large compressive 
  forces $\sim 2$\,kN at each spring
  as the barrel contracted to operating temperature, 
  which was shown in early modelling to be optically unacceptable.
 The selected combination of springs+rods in series provides 
   modest lens stresses and good lens centration. 
  Modelling also showed that 
   even in an unrealistic fast-cooling scenario with the rods at 293\,K and
  the lens barrel at 110\,K, the compressive forces remain well 
   inside the safe limit for Infrasil; therefore thermal shock during 
  camera cooldown cannot harm the lenses.  

 Also, each lens is thermally coupled to the barrel
  by copper flexi-straps which are held between the 
 PTFE rods and the lens edge. This ensures that the lenses
 are thermally coupled to the barrel at 12 points, which means that the
  lens temperatures are nearly axisymmetric inside the optical beam; 
 this minimises asymmetric temperature-dependent refractive index 
 variation. (Symmetric variation is present, but is essentially pure
 focus shift which is corrected by active optics).  

{\newtwo The lens barrel is designed to minimise stray light: 
 the lenses are generously oversized by $> 20$\,mm radius beyond
 the optical aperture, so all lens edges are shadowed from the window
  or hidden from the detectors. 
 The inward-facing surfaces of lens supports are V-grooved 
  to minimise single-scatter stray light, and coated with 
 Aeroglaze Z306 black paint. }

\subsection{Filter Wheel} 

The VIRCAM filter wheel rotates to place a selected
 passband filter in the optical beam: it is a large annular wheel of
 1.37\,m diameter and overall mass of 75\,kg, with its rotation
  axis offset 500\,mm from the optical axis.  
 The wheel has eight main positions equispaced at 45 degrees;  
   one is reserved for the Dark 
 (opaque) calibration filter, thus seven sets each of 16 individual 
 science filters are accommodated. 

The wheel is moved in open-loop using a commercial stepper motor 
 customised for cryogenic operation. The motor drives the wheel via
 a Delrin polymer worm: the gear ratio is 210:1 and the stepper motor
  uses 1000 half-steps per revolution. 
 The wheel can rotate in either direction and
 usually takes the shorter path,  but all moves
 are completed by moving to a motor position +1000 counts
 from the target, then moving 1000 counts in the negative 
  direction at low speed;  
  thus all moves are terminated in a standard direction 
  to minimise backlash errors.  
 
A datum switch is used to initialise the zero-point of the
  wheel step count.   
 To avoid a single-point failure, 
  a second (offset) datum switch is provided, which
 can be selected by re-plugging wires outside the cryostat and 
 telling the software that the backup datum switch is connected. 
 
Also, a third ``near-position'' switch is provided which 
 clicks on/off at every science position, 8 times per revolution. 
 This cannot discriminate which filter is in use, and
 is not used in closed loop:  but it provides
 a sanity-test against gross wheel errors, and logging
 the step count each time this switch changes state
 can provide information about wheel non-repeatability. 

 Each science ``filter'' is actually a mosaic of sixteen individual
  {\newtwo 54\,mm square} 
  glass or Infrasil panes, one per science detector, all mounted in an
  Al tray using spring-loaded retainers.  
 Currently the filter complement is Z, Y, J, H, $\Ks$, a
  narrow-band $1.18 \mic$, and a ``dual'' filter with
  narrow-bands $0.98 \mic$ and $0.99 \mic$ covering half the detectors
  each.  
  The complement of filters in the wheel can be exchanged
   (with the camera warm and off the telescope) 
  via a dedicated filter access hatch on the cryostat bulge. 
  The hatch is on the upper side of
  the wheel, away from the detectors, to minimise risk in  
  case of a stray part falling from a filter tray. 

 The filter wheel also contains eight smaller ``intermediate'' slots
 in the V-shaped gaps between the rectangular science filters. 
 Each intermediate slot covers only three science detectors; 
 currently one of these slots accommodates the two 
 high-order wavefront sensing beamsplitters (see Sect.~\ref{sec-actopt}), 
  and the remaining 7 are blanked off. 

The structure outside the filter wheel also includes
 baffling to minimise scattered light bypassing around the wheel edge
 and reaching the detectors. This is very successful; 
  the measured count rate with the dark filter is $< 10^{-5}$ 
 of that through the science filters, 
  so low that dark frames do not change if
 the dome lights are turned on.

\subsection{Cold Baffles}  

The cold baffle stack is a critical element of the Camera:
 it serves two functions, firstly to block unwanted heat
 radiation from reaching the detectors, and secondly
 to minimise radiative heat loss from the window into the cryostat.  

\begin{figure}[t!]
\centering
\includegraphics[width=9cm]{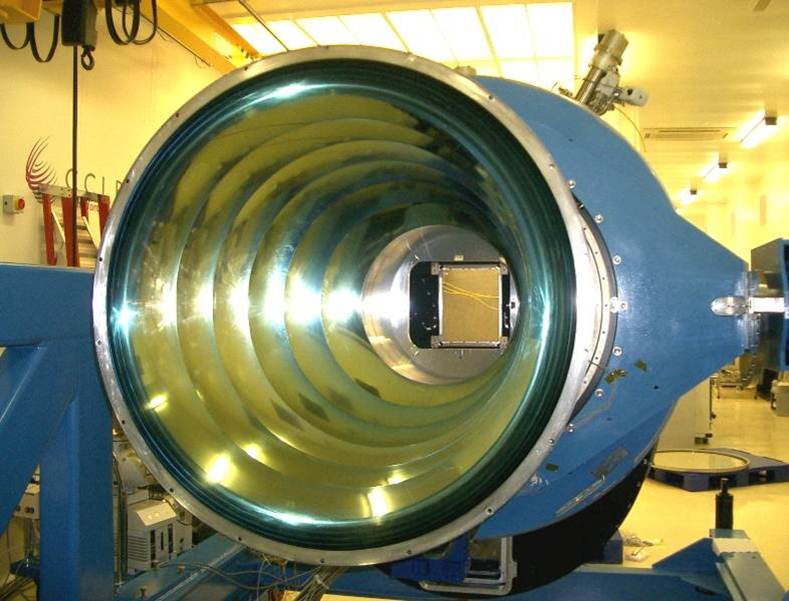}
 \caption{
 VIRCAM cold baffle, with the cryostat nose and window not present. 
 } 
\label{Fig-baffle} %
\end{figure}

 The 250\, kg cold baffle structure (Fig.~\ref{Fig-baffle}) 
  is made of Al alloy: the geometry
  is an outer support cylinder, with seven nested
 ellipsoidal baffles inside the cylinder, and one flat annulus on top. 
  The flat annulus with aperture diameter 812 mm defines the 
 entrance beam to the camera; the ellipsoids are sized
 so that they clear any ray from the top baffle to the detectors.  

 The baffle support cylinder is attached at the bottom end to the main 
  Optics bench: due to the substantial overall height 1.82\,m, 
   it has a significant temperature gradient,
  running from 100 K at the bottom to 160 K at the top, but the top
  end by design remains cold enough that its own thermal emission 
  contributes negligibly to the detector background. Keeping
  the baffle top acceptably cold required a substantial
  wall thickness for the baffle support cylinder, hence the
  substantial baffle mass. 

 A significant design challenge was to avoid overcooling
 of the window: the Infrasil window is opaque at thermal-IR 
  wavelengths $> 4 \mic$, 
  and therefore emits $\sim 100 $W of blackbody radiation
  downward into the cryostat. If most of this were absorbed,
 the centre of the window would chill $> 10\degc$ below ambient 
 temperature, leading to a severe problem with 
  dewing or frosting on the window's outer surface.  
 Since Infrasil is a poor heat conductor, 
  heating the window edge would not remedy this. 
 Therefore, most of the heat energy radiated by the window must
  be reflected back to the window. 

 The ellipsoids in the baffle are shaped so as to reflect 
  as much heat as possible back to the window. However,
 simple metallic heat reflectors would not be acceptable
  at $\Ks$ band, since emission from the dome 
 can reach the detectors via two reflections in the
  baffles and M2. As viewed from the detectors,
  the cold baffles reflected in M2 subtend 
  around 15\% of the total M2 solid angle, so fully-reflective
   baffles would give almost 15\% {\em additive} 
 emissivity contribution, clearly unacceptable. 

 The solution to these two opposing requirements uses the
  fact that for a room-temperature blackbody, almost all the radiated power 
   is emitted at long wavelengths $\lambda > 5\mic$, much longer than the
  science bands: therefore, 
  the adopted solution is to apply a {\em dichroic} 
 coating on the baffles which is a good reflector at
  $\lambda > 4 \mic$ but is strongly absorbing at science
 wavelengths $< 2.5 \mic$. This custom coating was developed specially
 for VIRCAM and applied to the upper surfaces of the
  ellipsoid baffles by Reynard Corp. of California.  
 
 All surfaces visible from the detectors, i.e. 
 the inner face of the cylinder, and the downward-facing
 sides of the ellipsoid baffles, are finished with
  Aeroglaze Z306 black paint for maximum absorption of 
 scattered light: also, the ellipsoid baffles are spaced so that
  no point on the window can see the inner face of
 the cylinder, therefore the only single-scatter
 paths from the window to the detectors are the sharp-edge tips of the
  ellipsoid baffles. 

\subsection{Detectors} 

The camera's ``retina'' comprises sixteen Raytheon VIRGO HgCdTe 
 infrared detectors (Love et al \cite{love}), 
 each of $2048^2$ format and $20 \mic$ pixel size. 
 For efficient tiling of sky, the detectors are arranged 
 in a $4 \times 4$ rectangular grid, 
 with spacing of $0.9\times$ the active width in the detectors $y-$direction 
 (which is cryostat $-x$) 
  and $0.425\times$ active width in the orthogonal direction.

 The detectors have an operating wavelength range (half-peak)
  $\approx 0.75 - 2.45 \mic$, and the quantum efficiency 
 is very high from $1.0$  to $2.35 \mic$; measured QE is near 90\% for
  most of the detectors across this range (Bezawada et al \cite{bezawada}). 
 At the short end there is a gradual QE rolloff
 to $\sim 75\%$ at $0.8 \mic$, and a sharp cutoff below
  $0.75 \mic$. 
 
 The total of 67 Mpixels makes this the largest near-IR focal plane 
 in astronomical use; a number of other astronomical 
 instruments have four $2048^2$
 detectors giving 16.8 Mpix, but VIRCAM is the only instrument 
  to exceed 17 Mpix at the present time.  

 The detectors comprise a CdZnTe ``substrate'' above the active layer,
 then the active HgCdTe layer; this is bump-bonded via
 4 million individual Indium bumps to the silicon {\em multiplexer}
  which reads out the signals. 
 The silicon is attached via epoxy glue to a thick Molybdenum
  back-plate: the Mo is a very good CTE match to CdZnTe, and
  thus forces the silicon to contract at the same rate as the rest of
  the detector during cooldown.   

 Readout noise is typically  $24 \, e^-$ rms for a single 
  integration with double-correlated sampling (well below
 sky noise for broadband filters);  
 dark current is typically $0.2 \, e^- \, {\rm s}^{-1}$, with a tail
 to larger values.  
 The detectors do have a significant number
 of cosmetic defects, with typically 1 percent dead or hot pixels;
 one detector has some large ``dead patches'' totalling $\sim 3$ percent
 of its area. 
 Most of the dead or hot pixels are time-invariant, so the
 effect largely disappears in post-processing of jittered frames;
 however, this does mean that it is advisable to observe at
 least 4 distinct jitter positions.  

\begin{figure}[t!]
\centering
\includegraphics[width=9cm]{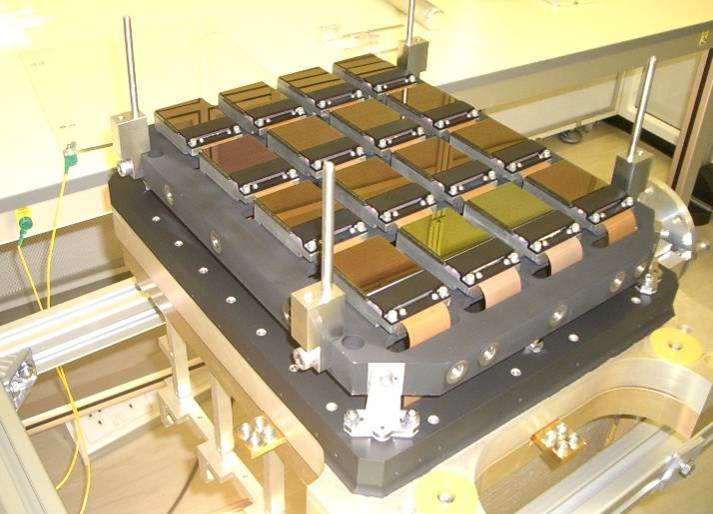}
 \caption{VIRCAM focal plane. The active areas of the 16 detectors 
  are seen with the bronze-coloured anti-reflection coating. 
 Black covers are over the detector electronics, and the ribbon 
  cables are seen leading to cold electronics below. One of the three
 knife-blade supports can be seen at the front corner. 
 The four vertical rods are temporary guides for a protective cover.   
 } 
\label{Fig-dets} %
\end{figure}

The 16 individual science detectors (Fig.~\ref{Fig-dets}) 
  are bolted to a thick Molybdenum 
  mounting plate (matching the detector back-plates) 
 which has a waffle-type lightweighting pattern on its back side,
 and includes feedthrough slots for the detector ribbon cables. 
 The plate was machined accurately flat, and detector 
 thicknesses matched so all pixels are coplanar within $\pm 12.5 \mic$. 
  The plate is held in position from 
  the Al support frame via a central boss providing centration, 
  and three titanium ``knife blades'' at the edge constraining
  rotation and tip/tilt.   The mounting arrangement allows for differential
  contraction between the Mo plate and the Al support structure
 with minimal stress on the Mo plate. 

 A cold electronics box attaches to the back of the Al detector support
  frame. If necessary, 
  the complete subsystem of detectors, Mo plate, support frame
 and electronics box can be removed via the 
 back of the Camera as a single $54\,$kg assembly, 
 after warming up the Camera and removing the cryostat back-plate
 and thermal shields.  

\subsection{Detector electronics and readout} 

The detectors are controlled and read out by ESO IRACE (InfraRed
 Array Control Electronics) controllers. One master box is used 
 for clock signals, driving four slave controllers each controlling
 a column of four detectors.  Each detector has sixteen parallel
  readout channels, each 128 pixels wide by 2048 pixels high,  
 thus overall there are 256 parallel readout channels. 

The default readout time is 1 sec (pixel rate 294 kHz per channel, 
 including reference pixels).  
 The standard readout mode is double-correlated sampling; 
 for a given detector integration time (DIT), 
  the detectors are reset, then a non-destructive read
 is immediately done, followed by a wait for DIT $-\; 1$\,sec and
  another read. 
(Windowed readout is possible for faster operation, but
  is not offered as a standard operating mode; this is 
 constrained to read the same sub-window on 
  all 16 detectors). 

The median system gain is 4.19 $e^-$/ADU, and
  readout noise is typically 24  electrons for a single
 integration with double-correlated sampling; this is 
 well below the sky noise for typical broadband exposures.  

Detector linearity and dark pixel calibration is done
 using the flat-field screen and calibration lamps 
 inside the enclosure (see Sect.~\ref{sec-data}). 
 Non-linearity is significant,
 typically 2--4 percent at 10,000 ADU flux level 
 with one detector at 10 percent, but is fairly stable
  over time and within each detector.  The non-linearity 
 is corrected in the data processing pipeline
  using polynomial fits to the observed linearity sequences. 

 Normal exposures are done by co-adding several individual 
 detector integrations in the IRACE electronics before saving to disk;
  (e.g. $6 \times 10$sec for $\Ks$ band, or $3 \times 20$sec for J).  
 this reduces data volume and overheads, since consecutive integrations 
 co-added in electronics have an overhead of 2 sec each, while
 the overhead in the IRACE for processing  
  a complete exposure and clearing buffers 
 is 4 sec; this sets the minimum time between consecutive 
 exposures, since the N+1th exposure cannot start until the
 Nth exposure has been cleared.  
 However, telescope jitter/offset moves
  are operated in parallel with the clearing of IRACE buffers, and
 these usually take longer than the 4 sec, so this rarely adds
  overhead.  


\subsection{Autoguider and Wavefront sensors} 

The VIRCAM contains six auxiliary CCDs (two units, 
 each with two WFS CCDs and one autoguider CCD) to provide
 autoguiding and wavefront sensing; however, since these
 are functionally part of the telescope active optics system rather
 than the Camera, we  postpone description of these until 
 the next Section \ref{sec-actopt}. 

\subsection{Camera optical testing} 

The Camera was challenging to test in the lab, due to the long
 cooldown/warmup time, and the fact that the Camera and Telescope
 form a linked optical system: thus, for image quality 
 testing we constructed a ``telescope simulator''
  which provided input beams with deliberate aberrations
 closely approximating those of the real telescope at 
 two specific field locations, ``near-axis'' at 30\,mm 
  and ``off-axis'' at 112\,mm (Dalton et al \cite{dalton10}).  

The telescope simulator comprised a small Cassegrain-like
 telescope with 0.6\,m spherical primary mirror and 0.2\,m 
 spherical secondary, together with small corrector lenses. 
 The corrector lenses included a cylinder lens to generate
 the required large off-axis astigmatism; by suitable re-spacing of these
  lenses, this simulator with a pinhole source 
  generated an $f/3.25$ converging beam into the Camera, with
 aberrations close to the real VISTA M1/M2 at either 0.14 or 0.53
 degrees off-axis.   

 For these tests, the Camera was built in a ``short nose'' configuration
 with the real lens barrel but without the cold baffle and using
 a smaller-diameter 60\,cm BK7 test window, 
 since the real Infrasil window (as expected) was not delivered 
  until the final stages of Camera testing.  All tests were done at 
 $J-$band only for simplicity. 

This camera optical testing procedure consumed a significant amount of
 time and effort, and due to the challenging tolerances
 for relative alignments of the test source, it did not quite meet the
 target accuracy level we were initially hoping for: however,  
 it did achieve the key purpose of giving an end-to-end independent
 proof that there was no severe error in the camera optics, 
 in advance of flying the camera to Chile.  

During commissioning on-sky, we found a moderate
 astigmatic term $\sim 300\, {\rm nm}$ which co-rotates
 with the Camera and is roughly constant across the field 
 of view; the cause is not known for certain, but is suspected to
 relate to a thermal gradient across lens L3  
 and consequent refractive index gradient. This was corrected
 quasi-automatically by the active optics system, 
 but to eliminate correction delay after a slew
 we added an active force term to the M1 in software which co-rotates
 with the Camera.

\subsection{Handling and transport} 

 A large purpose-built
  earthquake-resistant camera trolley was made, and this
 is used both for camera assembly at RAL Space
  and off-telescope maintenance at Paranal. The trolley includes
 pitch and roll bearings to re-orient the camera
  during assembly and maintenance operations. 

 For transport, it was decided to fly the 
 camera to Chile fully assembled including detectors. 
 A customised container
  was designed which takes the camera on the trolley. The complete
 package of Camera, trolley and container weighed over 9 tonnes,
 and was designed to (just) fit through the side
   door of a Boeing 747 cargo aircraft.  
 For safety, the trolley was linked to the container floor 
  via substantial wire-rope shock isolators;
  these were specified so that a computer-modelled 
  20\,cm free-fall accident should inflict
  no more than 5\,g peak deceleration on the Camera.  
 In reality, the journey was smooth
  and the camera shock recorders measured no shock 
 exceeding 0.5\,g during its flight
 to Santiago and subsequent road transport to Paranal. 
  
 Due to the long camera nose, 
  mounting the Camera onto the telescope is necessarily
 done with the telescope pinned horizon-pointing, using the
  main enclosure crane. Since the 
 camera centre of gravity is inside the Cassegrain
 rotator, this uses a purpose-made lifting arm which attaches 
 around the back face of the cryostat and has the lifting eye 3\,m above 
 the combined centre of gravity, above the M1 Cell.

\subsection{Camera shim} 
\label{sec-shim} 

 For adjustment of Camera position, 
 the design includes a substantial shim (made
 of nine sectors) between the
  instrument flange and the Cassegrain rotator;  
  this had a nominal thickness
  of 25\,mm with a possible range from $5 - 45\,$mm. The purpose
 was twofold: firstly, the wide range allowed potential spherical
 aberration compensation in the event of conic constant error 
 in either mirror, by a combination of axial shift of the 
  whole VIRCAM and refocussing M2. 
 In reality, the as-built spherical aberration of M1+M2 was very close
 to predictions; however the as-built 
 VIRCAM window was 1.5\,mm thinner than designed due to
  extra fine-polishing to remove surface defects, so the 
 baseline 25\,mm shim thickness was reduced by 0.5\,mm to 
  compensate the thinner window.
  Secondly, changing to a new shim with a slight wedge angle allowed
 correction of the detector plane orthogonal to the Cassegrain axis, 
 avoiding the risky operation of opening the cryostat
  and adjusting detector mounts.  
 This tilt adjustment {\em was} used: near the end of commissioning,
  analysis of science images indicated a small residual
  focal tilt (see Sect.~\ref{sec-actopt}), so a second shim 
  was made with the same mean thickness but a $150 \,\mu{\rm rad}$
 wedge angle; this was installed in early 2010 and 
 reduced focal tilt to a negligible value.

\subsection{Camera performance} 

 Apart from several minor teething problems during commissioning, mainly
  regarding electronics problems and inadequate initial glycol supply, 
 the VIRCAM has generally been very reliable since the end of 
 the commissioning period.
 
 The cold baffle has proved to be a successful design: 
 the observed $\Ks$-band background is in line with the 
  computer modelling, while  
 the window centre is estimated to be $\sim 4 \degc$ below ambient temperature. 
 Dewing of the window has occurred a few times to date during rapid
  increases in humidity,  
 but the window has been successfully cleaned by careful hand-washing
 without degradation of the coating.  

 The vacuum and thermal design has proved successful, and the
 Camera can remain cold for the 1-year intervals between cryocooler
 major overhauls.  There have been no significant outgassing or
 contamination problems, and 
 since the Camera departed from the UK there has never been any re-alignment
 required of lenses or detectors inside the Camera, and
 only one internal intervention to fix a 
  problem with the filter wheel.  
{\newtwo Sensitivity and image quality are described in 
 Sect.~\ref{sec-perf}. }

\section{Active Optics} 
\label{sec-actopt} 

VISTA has a fairly sophisticated guiding and active optics system, 
  which includes a total of six auxiliary CCD detectors 
 (inside the VIRCAM) totalling 24 Mpixels. 
 The purpose is to control the telescope guiding, and the 
 {\em position} of the secondary mirror (in 5 axes)
 and the {\em figure} of the primary mirror (in up to 18 eigenmodes)
 to maintain near-optimal image quality. 

 As a notable feature, VISTA is probably the first telescope
 ever to deliver
  closed-loop 5-axis collimation\footnote{Of the 6 
 rigid-body degrees of freedom, 
   small rotations around the z-axis are irrelevant, 
  so 5-axis control is required for complete collimation.}   
 of the 
 secondary mirror {\newtwo in parallel with observing}; 
  traditional telescopes only
 used an occasional focus shift, while most modern telescopes 
  control 3 degrees of freedom of M2 in closed loop: usually, focus
  and $x/y$ rotations around M2 centre of curvature 
 \footnote{ {\newtwo We note that the VLT does occasional 
  on-sky 5-axis collimation; 
   however sensing the 4th and 5th degrees of freedom 
  (M2 tilt around coma neutral point) 
  requires stepping the single guide/WFS probe to multiple locations 
  around the field edge, so the VLT cannot do closed-loop M2 tilt
   control during science observing. Due
  to the $f/15$ system and narrower field, the VLT is much less
 sensitive than VISTA to misalignment in these axes.}}. 

 The VISTA primary mirror is only moderately thin
 by modern standards: its diameter/thickness ratio is 24, 
 compared to values of 48 for the VLTs,  18 for the NTT 
  and 6--10 for pre-1980 4-metre class telescopes. 
 This is beneficial, since the M1 is flexible enough that
 the figure can be modified with moderate 
 active forces, but not so flexible that
 very high accuracy is required for the support forces.  

However, the very fast f/1 primary mirror 
 and fairly fast f/3.25 system focal ratio mean that
 the system is extremely sensitive to misalignment
 and despacing between the M1 and M2:  
 the image quality budget requires that the M1-M2
  spacing is correct to $\approx 3\mic$ (i.e. fractional
    accuracy 1 part per million), 
 centration to $\approx 20 \mic$, and tilt to a few arcsec. 
  It was felt unrealistic to maintain these numbers with purely 
 open-loop control, therefore active measurement and control of the
 M2 position is mandatory. The IR Camera includes
 a number of wavefront sensors and autoguider units to
 provide the required corrections. 

 We show in Sect.~\ref{sec-wfsloc} below that measurement of M2 collimation
 requires measurement of only the 
  low-order optical aberrations (defocus, 3rd
 order coma and 3rd order astigmatism), but
 these must be measured at 
 {\em two} locations widely spaced in the field, in order
 to disentangle the effect of M2 tilt around the coma neutral
 point. 

 Measuring the figure of the M1 is best done by a
  single sensor near-axis, but up to 18 bending modes 
  are required; since the altitude dependence 
  is relatively repeatable, this can run mainly
  with lookup-table control, using only occasional
  measurements, mainly to eliminate any long-term drift in
  the M1 actuator load cells.  

 Due to these differences in cadence and mode requirements, 
 the wavefront sensing functions were
  split between separate low-order and high-order wavefront
 sensors. 
 In detail, there are two autoguiders (AG), 
  two {\em low-order wavefront sensors} (LOWFS), and
  two {\em high-order wavefront sensors} (HOWFS) units; 
  the autoguiders and LOWFS are joined in two LOWFS+AG units
   with a common housing. 
 The low-order wavefront sensors are required
 to run quasi-continuously (around 1 cycle per minute) to provide
 closed-loop focus and collimation of M2 ; while the
 high-order sensor(s) run only ``on demand'', typically
 once every evening twilight and many times on 
  occasional engineering nights;
 the latter data are then fitted off-line
 to build the lookup table for M1 active forces. 

\subsection{The 5-axis collimation and sensor locations} 
\label{sec-wfsloc} 

It is well known that for a 2-mirror telescope, despace of the
 secondary mirror produces defocus, and decentre of the secondary
 mirror produces mainly 3rd order coma ($Z_7$ and $Z_8$ in our
 Zernike polynomial convention);  the coma difference
 from an ideal system is nearly independent of field angle,
  so a wavefront sensor anywhere in the field is sufficient.  
  Most modern 2-mirror telescopes correct the
 latter by rotating their secondary mirror 
  around its centre of curvature, which gives almost pure coma
  correction and negligible image shift.  
 
However, defocus and coma information can only correct 3 of the 5 M2 degrees
 of freedom: applying this correction leaves the
 M1 and M2 axes crossing at the {\em coma neutral point} which
 is typically close to the primary focus {\newtwo (for VISTA the coma neutral
 point is 1.05\,m 
 above the M2 pole)}, but leaves tilt of M2 around
 the coma neutral point unconstrained (McLeod \cite{mcleod},
 Noethe \& Guisard \cite{noethe}.)  These authors showed that
 rotation of the M2 around the coma neutral point adds 
 3rd order astigmatism which is bi-linear in M2 tilt and field angle  
 (therefore has negligible effect on-axis).  
  We ran a Zemax analysis to verify that essentially the 
  same behaviour is found for the VISTA system including 
   the VIRCAM field corrector, and the aberration differences
 (relative to a perfectly aligned system) are nearly achromatic. Thus, 
  off-axis sensor(s) are required for full collimation, but 
  the choice of wavelength is relatively unconstrained.   
 In principle it is possible to measure M2 tilt with only one off-axis
 sensor: however, with one sensor the M2 tilt is degenerate with astigmatic 
 figure error on the mirror(s), which produces an astigmatic term 
  nearly constant across the field of view; correcting 
 a real M2 tilt with M1 forces, or vice versa,  
  would double the astigmatism at a field point diametrically 
  opposite to the wavefront sensor, clearly a very undesirable result.  

 Thus, we decided to use {\em two} low-order wavefront sensors
 in VIRCAM, located at diametrically opposite points near the edge
 of the field of view; the LOWFS hardware is described below, along
  with the autoguiders which are co-mounted.  
 
\subsection{Low-order wavefront sensors and autoguiders} 

 The purpose of the low-order wavefront sensors (LOWFS) is 
  to deliver measurements 
 every few minutes to maintain correct position of the M2 in all
  5 degrees of freedom, while the autoguiders (AG) compensate
 for slow tracking drift e.g. flexure or thermal residuals
  from the open-loop pointing model.  
  These must operate in 
  parallel with science observing, but the sensors 
 must reside behind the corrector lenses inside the cryostat,
 so no moving probes were permitted. 
 Thus, the LOWFSs and AGs require a wide field of view, 
  sufficient to provide $> 99\%$ probability
  of finding a usable star  even at the Galactic poles,
  assuming a 45 sec exposure for the LOWFSs and 0.2 sec for
  the AG.  

 This is achieved by picking off
  patches of sky which are inside the circular corrector field but
 outside the rectangular science field, and using CCDs operating
 at far-red wavelengths for acceptable image quality and reduced
  sensitivity to moonlight. (Near IR sensors were ruled out for
  both cost and sky background reasons).  
  All CCDs are E2V 42-40 2048$^2$ deep-depletion devices for optimal
  red performance; there are six CCDs, one for each AG and two for
  each LOWFS.  

  Physically, one LOWFS and one AG are combined into a common housing
  with three CCDs to form a ``LOWFS-AG unit'' (mass 2.2\,kg),  
  and two identical LOWFS-AG units are used;  each 
  LOWFS-AG unit includes a housing containing a pickoff mirror, 
  filter, beamsplitter cuboid, 
  one autoguider CCD and two LOWFS CCDs, and a heater element
  to warm the CCDs to $\sim$ 150\,K.  
 More details of the hardware are given by  Clark et al (\cite{clark}).   
  
  The two LOWFS-AG units are mounted 180 degrees apart 
  on an aluminium plate (with a hole 
  for the science beam) just above the filter wheel. 
  The VIRCAM cold structure is designed so that the LOWFS-AG mounting plate
  is bolted to the IR detector support frame via three thick pillars
  in a near-equilateral triangle; 
  one pillar goes through the hole in the centre of the filter wheel,   
  while the other two are outside the wheel.  
 This geometry provides negligible differential flexure between the
  LOWFS-AG units and the science detectors, as required.    

 After the pickoff mirror, the beam into the 
  LOWFS-AG unit passes through a fixed filter 
  (bandpass $720 - 920$\, nm), then  
  through a cuboid fused-silica beamsplitter (these elements
  are common to the LOWFS and AG detectors).  
 The  beamsplitter coating changes along the long axis of 
   the cuboid: one end divides the light
  equally between the pair of LOWFS CCDs, while the other end
  is fully reflective and feeds the autoguider CCD (the extra
   reflection just helps packaging). 
 The two LOWFS detectors are mounted $\pm 1\mm $ respectively before/after 
 the nominal best focus location, so all stars image into small
 doughnuts defocused to a diameter of $0.3\mm$ or 5 arcsec. 
 (In principle measuring only one side of focus is sufficient, e.g. Tokovinin
  \& Heathcote (\cite{donut}), but our implementation measuring 
 both sides of focus is generally more robust and provides
  reduced cross-talk if multiple aberrations are present).  
 The autoguider CCD is physically identical to the LOWFS CCDs, but operated 
 in frame-transfer mode with half the area used as storage to 
 increase the duty cycle; maximum rate is around 5 frames/sec, though 
  usually 1 - 2 frames/sec is selected. 

 With a pixel size of $13.5 \mic$ or $0.229$ arcsec, each LOWFS
 unit views a square $7.8 \times 7.8$ arcmin of sky; this is sufficient
 to almost guarantee finding a usable star ($I \simlt 15.5$) 
  at any telescope pointing, even near a Galactic pole.   
 The LOWFS (and AG) detectors have no shutter,
  but the LOWFS readout is windowed down to
   typically a $100\times 100$ box around the selected star; this 
 reads out in $\ll 1$ sec, fast enough that image trailing 
  is barely detectable.   

 Normally only one autoguider is operational, whichever one has a brighter
 star; the use of two autoguiders gives identical LOWFS/AG units, 
  wider sky coverage and partial redundancy against one AG failure. 
  Both AGs can be run simultaneously in engineering mode. 
 (In principle, two AGs could give closed-loop control of the Cassegrain
 rotator; however this is unnecessary given the very good
  open-loop performance of the Cassegrain rotator and rather short 
  single VIRCAM exposures, and the control software does not allow 
  this at present).  

 Care was taken to minimise adverse electromagnetic interference
  and ghosting effects from the LOWFS/AG units
 to the science detectors: their placement above the filter wheel 
  is well shielded from the science detectors, and the housing
  near the science beam has a sawtooth face to minimise scattered light.  
  The LOWFS/AG wiring exits the cryostat radially, while the 
  science detector wiring exits axially at the back-end 
  to maximise separation.
 Each LOWFS/AG filter comprises 3\,mm of Schott RG9 glass 
 (transmitting 720 - 1050nm) with a short-pass
  920\,nm coating on the second surface to reduce sky background. 
   The RG9 is strongly absorbing at $\lambda > 1.1 \mic$ so 
  J,H,$\Ks$ bands are reflected by the coating but absorbed 
  by a double-pass through the glass; Z-band of course is absorbed
  by the CCDs. 
  The LOWFS filter thus behaves as a retro-reflector around Y band, 
  but this has not shown significant ghost problems so far. 

\subsection{High-order wavefront sensors} 

The high-order wavefront sensing only runs ``on demand'' by
 re-pointing the telescope at a suitable bright star. For simplicity, 
  a novel beam-splitter cuboid was designed which fits into
  unused space in the filter wheel (in one of the V-shaped 
 gaps between science filters), and produces a pair of offset
 above/below focus images on a
 science detector.  

 The beam-splitter comprises a cemented cuboid with two 
 tilted plano part-reflective surfaces immersed inside it. 
 The overall thickness is $\approx 6\mm$ thicker than a science
 filter; this extra silica produces a focus shift so the
   straight-through image has a best-focus about $2\mm$ 
 {\em below} the science detector. The spacing of
 the part-reflective surfaces is chosen so the doubly-reflected path 
 produces an image which has best-focus $2\mm$ {\em above} the science
 detector, and is also offset laterally by 100 pixels.  
 The upper face of the cuboid has a regular J-band coating. 

 Thus, to operate the high-order wavefront sensor, the filter
 wheel is rotated to centre the cuboid above a selected pixel
 on the science array, and the telescope is re-pointed
 to put this star close to the selected pixel; then
 a 60 sec exposure is taken to average out atmospheric
  variations.   This produces
 a pair of offset doughnut images on the science detector, 
 which are then analysed (see below) to measure the wavefront aberrations. 

In practice, we supplied two identical cuboids which are 
  mounted at different radii 
 from the axis of the filter wheel; 
 as the wheel rotates, each of these travels an arc 
  crossing four science detectors, 
 so by a suitable choice of wheel angle and telescope pointing, 
  any one of eight defined spots on different science detectors 
  can be used for high-order
 sensing. Normally a default spot near-axis is used, but all eight
  accessible spots were used occasionally during commissioning. 

 This high-order WFS is a simple and elegant solution which 
  uses no additional moving parts, 
 no additional detectors, while delivering good coverage of
  field positions. No motion of the secondary mirror is required --
  the observed pre/post focus offsets are determined purely by the
   dimensions of the beamsplitter cuboid, so as long as the cuboid is 
  manufactured correctly it does not require precise location 
  in any direction. 

 \subsection{Image analysis} 

Both wavefront sensors use a common image-analysis package
 to analyse a pair of above/below focus doughnut images 
 and estimate wavefront coefficients in various modes. 
In summary, a subroutine is provided which 
  given any vector of Zernike coefficients:
 simulates a pair of  doughnut images  using the known
 defocus distances and system $f-$ratio, 
  using geometrical optics ray-shooting and Gaussian blurring 
  for seeing; 
 then, a Nelder-Mead simplex algorithm searches 
  the many-dimensional space of Zernike coefficients
 for a minimum $\chi^2$ between observed and modelled images.  

 This ``forward'' search 
  algorithm is relatively slow compared to ``backwards''
 methods which attempt to reconstruct a wavefront directly from the
  data, but it is robust against fairly bad aberrations,
 and generally converges in $\sim 15$ sec for the low-order WFS
 and $\sim 500$ sec for the high-order WFS (with many more modes). 
 These time delays are acceptable since the low-order WFS runs in
 parallel with observing, while the high-order WFS usually runs in 
 bright twilight before science observing starts.

\subsection{Applied corrections} 

The degrees of freedom controlled by the active optics 
 system are the position of M2 in five axes, 
 and the shape of M1 in 17 degrees of freedom 
 (counting non-axisymmetric modes
  as two, with independent $\cos m\phi$ and 
  $\sin m \phi$ components).
 Here the mode $B_{m,1}$ is defined to be the
 softest mirror mode of azimuthal symmetry $m$; that is, the 
 force pattern of the form $F_i \cos(m\phi)$ where 
  $F_i$ for $(i = 1 \ldots 4)$ are arbitrary forces
 per ring;  which produces
  the largest ratio of rms deflection per unit rms force. 
 The $B_{m,2}$ mode is the next softest, orthogonal to $B_{m,1}$. 
 In order of increasing stiffness,  
  the M1 bending modes used are as follows: 
\begin{enumerate} 
 \item $B_{2,1}$ is the M1 lowest-order mode of 
  symmetry $m = 2$ (very similar to Zernike astigmatism).
\item $B_{3,1}$ is the lowest-order M1 $m = 3$ trefoil-like bending mode. 
\item $B_{4,1}$ is the lowest-order M1 $m = 4$ quatrefoil-like mode. 
\item $B_{2,2}$ is the M1 next-to-leading order mode of azimuthal symmetry 2. 
\item $B_{5,1}$ is the lowest order $m = 5$ pentafoil-like mode. 
\item $B_{6,1}$ is the lowest order hexafoil mode.  
\item $B_{3,2}$ is the next-to-leading order $m=3$ bending mode. 
\item $Z_{11}$ is Zernike spherical aberration. 
\item $B_{1,2}$ is an M1 next-to-leading order coma-like mode. 
 \end{enumerate}  
 Two other possible M1 modes are not used: 
  the $B_{0,1}$ mode is very similar to Zernike defocus,
  and the $B_{1,1}$ mode is very similar
 to Zernike coma. These modes are not applied as M1 forces,  
  but are corrected by the M2 only, to avoid a possible runaway 
   situation with opposing M1 and M2 corrections.  
 All the above modes and matching force 
 patterns were derived from finite-element modelling
 of the primary mirror using the pre-defined positions of
 the axial supports. 

In normal operation, the five M2 axes 
 and just {\em two} M1 modes (the pair of $B_{2,1}$ astigmatic modes) 
  are controlled in  closed-loop, while all higher
 M1 modes are controlled in open-loop via the lookup table, 
 including a once-nightly
 additive correction from the most recent high order wavefront sensor
  measurement. 

 The mapping from wavefront sensor measurements to 
 applied corrections is slightly complicated, and
 is essentially applied in three ``layers''. 
  For operator convenience,  all tables and corrections are stored
  and displayed as delta-wavefront per mode in nanometres rms,
  and then converted into M2 position offsets or M1 force patterns
   when sent to the hardware. 
 
 The first layer comprises a lookup table, actually
  a set of polynomial fit coefficients, one
  for every mode above. In principle the software 
 can handle up to 4th order Legendre polynomials in Altitude
  and a linear term in temperature; but in practice
 a constant + linear Altitude dependence is a good enough fit 
 for most modes, with quadratic terms only for defocus and one other. 

The second layer is another additive constant per mode, measured
 from the high-order wavefront sensor. 
 The third layer is the closed-loop cumulative correction 
 derived from the low-order wavefront sensors. 
 The sum of all three layers is actually sent to the system, but storing
 the layers separately has two advantages: 
 firstly the lookup table updates immediately after a slew 
  (keeping the previous additive LOWFS correction), 
 so for moderate altitude changes $\simlt 15\,{\rm deg}$, observing
 can resume immediately without
  waiting for a new LOWFS correction; secondly, 
  this setup ensures a simple un-do operation 
 in the event of a bad wavefront measurement. 

For the closed-loop measurements, the average of the
 two low-order wavefront sensors is used to give
 focus $Z_4$ and coma $Z_7, Z_8$ corrections to M2 position, 
  and the two $B_{2,1}$ M1 force corrections.  
The {\em difference} of the two astigmatism measurements 
 between the two wavefront sensors 
 (suitably rotated) is used to derive M2 tilt correction
 (rotation around the coma neutral point).   

\subsection{M1 / Camera Collimation and focus gradients} 
\label{sec-focgrad} 

The closed-loop procedure above is necessary and
 sufficient to align the M2 to a position which gives
 the as-predicted coma and astigmatism for
  any given M1 and Camera position: 
 if the M1 and Camera are almost co-axial, this is
 sufficient to collimate M2 to the same axis and 
 the collimation is essentially ideal.  

 However, if the M1 and Camera axes are misaligned to each other, 
  there is generally no M2 position which 
 can eliminate focus gradient and field-dependent 
 astigmatism at the same time: 
 centring M2 around its centre of curvature will still eliminate
  excess coma,  but to correct either focus gradient or field-dependent 
  astigmatism 
  one must then rotate M2 around the coma neutral point:  
 and the required M2 tilt vectors are different so 
  one can eliminate either astigmatism or focus gradient, 
   but {\em not} both. 
 
 The procedure is also complicated for an Altazimuth system:  
  since the Camera rotates about the $z-$axis but M1 does not, 
  it can be seen that {\em both} the M1
  and Camera must be separately aligned to the Cassegrain rotator axis
 to deliver a well aligned system at any sky position; and it is the
 M1 hyperboloid axis which is relevant, not the centre of the hole, 
  so this can only be determined by on-sky observations.  

 Our LOWFS system is set up to move M2 to deliver astigmatism matching
  its ideal well-aligned value (not zero), 
  because measuring a general focus gradient in one shot 
   requires 3 wavefront sensors, not 2;  
  thus, if the M1 and Camera are not well aligned, 
  a residual focus gradient will remain after active optics correction
   of the M2 position; this is near zero on-axis due to the symmetry of
 the LOWFSs.  
  This focus gradient will decompose into two distinct $x,y$ vectors: 
  one vector fixed to the M1 Cell (caused by fixed misalignment
 between the M1 axis and the rotator axis), and another
   vector co-rotating with the Camera (caused by misalignment
   between the Camera axis and the rotator axis). 
 Since the LOWFSs 
  also co-rotate with the Camera, the M1 misalignment term is observed
  as a focus difference between the two LOWFSs 
  with sinusoidal dependence on rotator angle.   
 The misaligment term co-rotating with the Camera cannot be measured
  by the LOWFSs (one component is invisible, and the
  other component is visible but 
  degenerate with static height difference 
  between the two wavefront sensors).  
 
 Both of these gradient terms were diagnosed and eliminated 
 during commissioning. For the M1-rotator alignment, we 
  used a specific on-sky  
  procedure: tracking several stars past the zenith, starting 
 from different Cassegrain rotator angles, while logging
  wavefront sensor readings, creates a plot of focus difference between the two 
  LOWFSs against Cassegrain rotator angle: misaligment 
  creates a sinusoidal dependence as above. From the amplitude
  and phase of this,  the required M1 lateral shift can be computed,
  and applied manually via the lateral definer adjustment screws.   

 For the Camera term we used data from the science detectors: 
   blinking pairs of images with M2 offset above/below best focus 
  at several Cassegrain rotator angles was used
  to estimate the component of focus gradient co-rotating with
  the Camera. 
  This tilt estimate was used to manufacture a new shim-ring between
  Camera and Cassegrain rotator, with a 
  wedge angle $\approx 150 \,\mu {\rm rad}$ to correct the 
  measured gradient.  

 Both of the above procedures took some time to develop and 
  optimise; an added challenge occurred 
 because a documentation inconsistency led the LOWFS software
  to initially contain a systematic focus error 
  (dependent on the off-axis distance of the LOWFS star 
   within the CCD). This error
  was of an annoying size, i.e. large enough to be significant but 
  not large enough to be readily obvious. 
 This created an apparent ``phantom wobble'' $\sim \pm 150$\,nm 
  in focus measurements only, which confused the interpretation 
   until it was eventually diagnosed and then easily fixed.  
 After this fix, the procedure worked very well; focus gradient 
  became negligible at all Cassegrain rotator angles, 
 and has remained so.  
 The mechanical structure between M1 and Cassegrain rotator is very rigid 
  by design (Sect.~\ref{sec-tel}), 
  so that this one-time alignment of M1 and Camera is sufficient. 
  A re-run of the M1 alignment procedure may be
   required after recoating of the primary mirror.  

\subsection{Alignment bullseye} 

It is instructive to count degrees of freedom (d.o.f) in alignments: 
 defining the Cassegrain axis as the datum and ignoring $z-$rotations,
 there are five d.o.f each for M1, M2 and Camera, 
 minus one for common-mode $z-$translations, hence $15-1 = 14$ d.o.f.  
 Two of these are M1-Camera axial spacing (set by metrology) and M1-M2 
 spacing  (closed-loop focus); the 
  symmetry of the LOWFSs decouples these
  from the 12 d.o.f's of tilt and decentre of M1, M2 and Camera. 
 In VISTA the 4 M2 tilt/decentre d.o.f's are controlled in closed loop
 as described above, while we did one-time corrections of M1 decentre 
  and Camera tilt for another 2+2 d.o.f; this leaves the final 4 d.o.f.'s  
 un-corrected. 

A Zemax analysis revealed an interesting explanation: 
 the analysis showed that there is a ``target plane''
  at $z = 1.02$\,m above the M1 pole, and {\em if} the combined 
 small tilts and decenters make the 
  M1 and Camera optical axes intersect (at a small non-zero angle) 
 in this target plane, then 
 the resulting 4-d.o.f. M2 compensation derived from the LOWFSs 
 leaves the images almost identical 
 to a fully-aligned system;   
  coma and astigmatism are set to ideal values by the M2 compensation, and
 the focal gradient is also negligible for this specific combination. 
 In practice, the axes do not have to cross exactly but 
  it is sufficient for the M1 and Camera axes to cross this
 plane within $< 0.2\,$mm of each other.  

To ensure this condition holds at any Cassegrain rotator angle, 
 we may then define a virtual ``bullseye'' of $\approx 0.1\,$mm radius 
 centred on the rotator axis at 1.02\,m above M1: 
 {\newtwo the adjustments to M1 centering and Camera tilt
  described above in Sect.~\ref{sec-focgrad} effectively align the  
   system such that {\em both} the M1 and Camera axes 
  pass through this bullseye}; it is then readily seen that the
 above condition will be satisfied at any rotator angle, and 
 the active optics control of M2 will deliver near-optimal images. 
 The 4 uncorrected d.o.f's then 
  correspond to small rotations of M1 and Camera 
 around the bullseye centre, which after M2 compensation 
 do not degrade image quality at first order.   
 (For gross combinations of tilt/decenter, second-order
   effects will become significant; but in practice 
 the initial mechanical setup is good enough that tilt angles
 are $\simlt 1$\,mrad, so it is safe to ignore effects
  second-order in tilt angles). 
 
This condition essentially 
 corresponds to the ``subspace of benign misalignment'' 
in the terminology of Schechter \& Levinson (\cite{schect}), 
 and the bullseye above provides a convenient 
 geometrical interpretation for the VIRCAM case.  

\subsection{Wavefront sensors performance} 

 Overall, the VIRCAM wavefront sensing
 system is fairly complicated, particularly with respect to 
 software; this consumed dramatically more manpower 
 than initially estimated in developing, testing, debugging, and modifying 
  for robustness on the real sky. 
 
 On the positive side, the wavefront sensing has worked well 
 since commissioning; no faults have occurred on the cold side. 
 The system is generally reliable apart from occasional software glitches
 or freezes,  and imposes relatively little overhead on science observations.  
 There is no detectable interference between CCDs and science detectors. 
 Some interference is seen between LOWFS and AG CCDs in
  the same unit, but only rarely when the readouts 
   coincide by chance; this results in occasional rejected AG or 
 LOWFS frames,  but no loss of time.   

\section{Enclosure and infrastructure} 
\label{sec-encl} 

 VISTA is sited on its own subsidiary summit 
 at 1.5\,km NNE from the main Paranal 
 summit. The mountaintop was levelled to create the platform
 at 2518\,m elevation, 
  and a new single-track road was constructed, branching off
 from the main VLT road. 

The enclosure serves a number of purposes, including :
 \begin{enumerate} 
 \item To protect the telescope from adverse weather, rain and
  storms. 
 \item During observing, to mitigate telescope windshake in high winds,  
  while providing good ventilation and minimal dome seeing in 
   normal wind conditions. 
\item To reduce unwanted straylight e.g. moonlight on the
   telescope optics.  
\item To control the telescope and structures to nighttime 
  temperatures during the day. 
\item To deliver a constant-brightness light source 
   and screen for daytime instrument calibrations.  
 \item To provide convenient access to the 
   telescope during maintenance and handling operations. 
\item To house the various services, including computers, electrical 
  supply hardware, compressors, glycol pumps etc.  
\end{enumerate} 
 
 The VISTA enclosure detailed
 design and construction was managed by EIE, Venice-Mestre. 
 It is a relatively conventional design 
  following modern practice for minimising local dome seeing, 
  using a combination of active daytime cooling and  
   good natural ventilation during observing.  
 The dome can rotate fully independent of the telescope 
  pointing; this is required for certain lifting operations, and
 is also convenient for routine daytime maintenance and testing.  

\begin{figure}[t!]
\centering
\includegraphics[width=9cm]{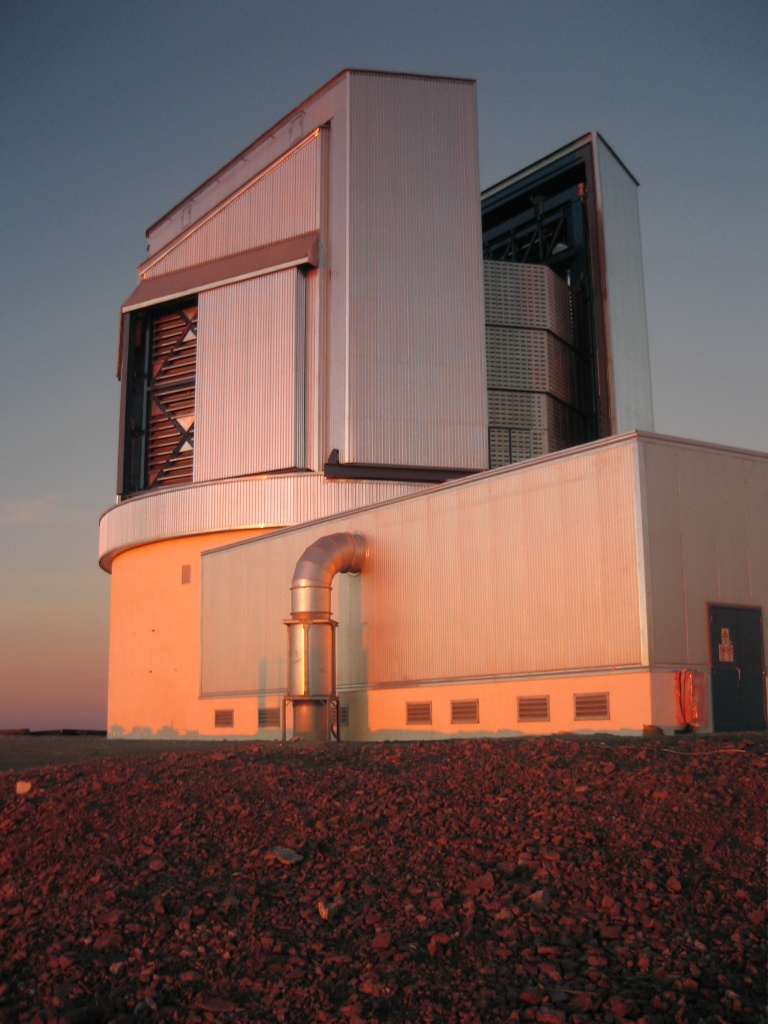}
 \caption{
  VISTA enclosure seen from the southwest at sunset,
  with the auxiliary building in the foreground. 
 Here the windscreen is fully raised, and an open ventilation door
  is seen on the left. 
 } 
\label{Fig-encl} %
\end{figure}

 The enclosure consists of three main parts (Fig.~\ref{Fig-encl}) :
  the rotating octagonal ``dome'' housing the telescope 
 is supported on a cylindrical concrete ``base''.  
 A single-storey ``auxiliary building'' adjacent to the enclosure base 
 houses the mirror coating plant, electrical supply rooms and
  a pump room for glycol and compressed air.  

\subsection{Enclosure base} 
 The enclosure base is a concrete cylinder of
  18\,m internal diameter, wall thickness 0.6\,m and height 
 6.2\,m, which supports the circular dome-rotation rail on its top face.  
 The ground floor contains six rooms around 2/3 of the circumference:
  there are two large rooms for the local control office and the Instrument
   Lab respectively, and four smaller rooms for telescope 
  electronics, Helium compressors, and a clean room if
 necessary to work on the detectors or electronics.  

 On the ground floor, a clear area of 6\,m diameter South of the
  telescope pier, next to the auxiliary building, 
 is used for M1 washing and all major lifting operations;
 a roll-up door connects this area to the coating
   plant in the auxiliary building. 
 An entrance lobby with a 3.1\,m wide door provides entrance
 for the Camera and other large items, though all planned 
  Camera maintenance can be done in the instrument lab 
  on the enclosure ground floor.  
 
 The enclosure base includes a concrete intermediate floor at height 4.6\,m; 
  this area houses the four main dome air-cooling cabinets, and 
  the slip-rings feeding power and control signals to the rotating 
  dome.  This floor also provides easy access to the telescope
  azimuth drives and the dome rotation mechanisms. 
  This intermediate air volume is actively cooled 
  both day and night, to provide 
 a thermal buffer zone between the warm ground-level rooms below  
 (typically $16 \degc$) and the telescope floor above.

\subsection{Dome} 
  The ``dome'' is the rotating structure housing the telescope; it is 
   based on a regular octagon in plan,
   and is a steel skeleton clad with insulating panels.
  All parts are straight beams and planar panels for
   cost minimisation. 
  The dome surfaces are clad with {\newtwo 80\,mm thick} insulating panels with 
  a bare aluminium external skin, to minimise heat load on the
  air-cooling systems during the daytime and to minimise 
  radiative over-cooling of the dome 
 exterior to the cold sky at night.  

  The dome is built on a large octagonal steel box-beam, 
  with a rotation bogie below each corner rolling
  on the 18.6\,m diameter circular rail; a non-load-bearing 
  circular skirt outside the octagon covers the gaps to the circular 
  concrete base. 
  The dome structure provides an
  internal clear diameter of $\approx$ 15\,m. 
  Four of the bogies are motor-driven for dome rotation, and
  the other four are idle. Each bogie includes a single 60\,cm
  main wheel, with conical surface to avoid skidding on 
 the circular rail; four small guide-rollers maintain centration
  on the rail. The dome can operate normally with any three motors 
  functional.  Power and control signals are supplied by slip rings
  above the intermediate floor level, so the dome has
  unlimited rotation in either direction. 

 The dome includes a 5.4\,m wide observing slit, 
 centred at a corner of the octagon, which
  is closed by two inverted-L sliding doors; these
  slide parallel to two faces of the octagonal dome to avoid
  protruding ``sails'' when open. 
 For earthquake resistance, the slit doors are pinned
  to the main structure in both the open and closed
  positions, by three motorised locking pins on each side of the slit.  
 When the slit doors are closed, a seventh locking pin joins
  the two doors at the apex to resist gapping in extreme wind 
  conditions. 
  Adjoining the top of the slit are two large trusses
  which support the dome roof and the main enclosure crane 
 (Fig.~\ref{FigTel}). 

  Six large ventilation doors (in three pairs) 
  are placed on the other six sides of the octagon:
  each pair of doors provides approximately $5 \times 6$m open
  aperture, for generous natural venting of the dome in low
  to moderate winds. 
 Fixed slanted louvres (see Fig.~\ref{FigTel}) 
 are bolted to the dome skeleton inside the
 ventilation doors, to allow good ventilation while rejecting
  unwanted stray light from the Moon.  

 A movable windscreen is located in the lower part of the observing
 slit: this is a 3-panel structure with the bottom panel fixed,
 and 2 overlapping panels sliding upwards; 
 the windscreen total height is adjustable
  from approximately 3\,m to 7.5\,m above the dome floor. 
 The windscreen is porous, with fixed open slots
  covering about 20\% of its area; this is
 supposed to improve ventilation when raised, and reduce
 vortex shedding from the upper edge. 
  
 The dome includes a deployable `moonscreen' which slides
  horizontally along the upper part of the slit, above
 the main crane.   In the retracted position this is outside the
  telescope beam to zenith, 
   while it can deploy forwards to fully cover a 4.5m circle
  centred above the telescope. This moonscreen 
  has several purposes:
 \begin{enumerate} 
 \item When appropriate, to reduce illumination of
 the telescope optics by the Moon. 
 \item To provide a calibration screen for daytime measurements:
   the main use is for linearity calibration which requires 
  a constant brightness source. 
 \item To protect the telescope optics from falling dirt or loose
  parts during opening/closing of the slit doors. 
\item To act as a sunshade if maintenance requires 
   opening the slit doors during daytime. 
\end{enumerate} 

 The ``observing floor'' (see Figure~\ref{FigTel}) is a flat 
  surface at 7.9\,m above ground level, made of three sections:
  an inner disk is attached to the Telescope yoke;
  an annular fixed floor (3\,m wide) is supported 
  on steel columns from the intermediate floor, and includes the two
  staircases;  an outer floor ring co-rotates 
  with the Dome. 
 Except for a fixed parking cradle for the telescope 
  top-end, the observing floor is not required to carry major loads, 
  so it is made of plywood over steel frame for low mass 
 and good thermal insulation. A conductive linoleum covering 
  provides protection against electrostatic discharge 
 (which is significant given the very dry environment at Paranal).  
 
 One unusual feature of the Dome is the placement of the rotation 
  rail and bogies {\em below} the observing floor level. 
  This somewhat increases the height of the rotating Dome structure, 
   but has several beneficial results:
 \begin{enumerate} 
 \item
  There is no concrete to form
   a heat-sink above the observing floor level.
\item 
  The slit doors and ventilation doors
  have their bases level with the observing
  floor, giving optimal ventilation and convenient maintenance. 
\item
  The dome rotation motors are located in the intermediate floor
   region below the observing floor, and are simply air-cooled.  
 (The intermediate floor air volume is actively cooled  
   both day and night, so waste heat does not accumulate there). 
\item 
 The dome electronics cabinets on the walls also dump their heat 
  down to the intermediate floor volume, using fans which pull in 
  air from the dome and exhaust downwards through vents 
  in the rotating dome floor.  
  Thus, no glycol supply is needed to the rotating dome. 
\item 
 The placement of the rotating/fixed interface
 below the telescope floor means that the dome light-tightness
 is very good, so linearity calibration sequences
 can be taken during the daytime 
  without interference from sunlight leaks. 
\end{enumerate} 

 For daytime chilling of the dome and telescope, 
  four large vertical air-ducts are mounted on the 
 dome walls (see Fig.~\ref{FigTel}): 
 during the daytime, the dome is normally 
 parked with the slit South-facing,  where 
  these air-ducts line up with the four glycol-fed air-cooling
  cabinets underneath on the intermediate floor. 
  An inflatable seal closes the gap 
 between the dome and the base to 
 minimise air leakage, and 
  three ceiling fans blow air downwards to minimise 
  any vertical temperature gradient.  

 A 10-tonne capacity bridge crane is mounted from 
  the trusses on either side of the dome slit: this crane is
 used for all major lifting operations, including M1, M2  
  and Camera lifting.  The crane can 
 traverse $\pm2.2\,$m sideways and $\pm 6.5\,$m along the slit,
  including crossing above the telescope.  Large items
  (e.g. the Camera) are carried around the telescope by slowly 
 rotating the dome with the load suspended.  

The dome also includes internal catwalks at the rear, 
 along the sides of the slit for access to the crane, 
 and ladders for access to the flat roof and upper 
  slit door motors.  
 A battery-powered scissors-lift platform is provided to 
  reach equipment on the enclosure walls. 

\subsection{Auxiliary Building} 

 The Auxiliary building is a single-storey building,
 approximately 19\,m by 14\,m, which adjoins the Enclosure base
 on the South side.  This building houses
 various electrical and glycol services, and the mirror Coating Plant. 

 Incoming electrical power arrives as a 10 kV supply 
  via underground cable from the  
 Paranal central generators at base camp: a transformer 
  converts this into 3-phase 400\,V, then 
 a switchboard room distributes this around the
  building to the various subsystems. 
  An uninterruptable power supply (UPS) 
   battery bank supplies the computers and electronics:
  this UPS does not supply the telescope or dome drives, 
  but provides enough capacity to power
  all the electronics and computers for $\sim 30$ minutes,  
  enabling a graceful shutdown in the event of site power 
  failure (which is rare, but does happen once or twice per year).  

 A large ``pump room'' houses many pumps circulating 
  chilled glycol (as 33\% glycol/water solution) 
 to the various subsystems, and an air compressor.   
  A main glycol pump circulates the glycol from  
 a 1500 litre cold-tank in the pump room to 
 a 2-head chiller unit, located on the mountainside 80\,m SSW (usually 
  downwind) from the telescope.  The chiller unit runs to maintain
  the cold-tank between 8 to 10$\degc$ below the 
  {\newtwo internal air temperature setpoint},
  unless the dewpoint rises above that. 
  Four smaller pumps circulate the glycol from the cold-tank to
  the services: two pumps feed the 4 dome air-cooling
  cabinets, one feeds the telescope motors, electronics and Camera, 
   and one feeds the Helium compressors and the ground floor 
   air-cooling unit.   
  All these glycol pumps are actually
   in parallel redundant pairs, with changeover valves 
   enabling continuous operation.
  This feature is important, since the 4 Helium compressors
  feeding the VIRCAM consume $\sim 5\,$kW electrical power each, and
  require continuous glycol cooling. If their glycol circulation
  stops, these compressors rapidly trip themselves off 
  and the camera starts to passively warm up; this  
  is not unsafe but  interferes with observations or calibrations.

\subsection{Coating Plant} 

The Auxiliary building also houses the Coating Plant 
 which is used for coating both VISTA mirrors, and (if desired) can 
 also coat the secondary mirrors from the VLT and Auxiliary 
  Telescopes.  
 The coating plant was manufactured by Stainless Metalcraft (Chatteris) 
 and comprises a stainless-steel vacuum vessel 
 containing 3 (later 4) fixed magnetrons and a motorised mirror
 rotation system.
 The vacuum vessel comprises a spheroidal lower section, which 
 is mounted on rails and movable horizontally into the enclosure wash
 area; and a conical upper section, which lifts vertically on four screw-jacks 
 to open and close the vessel. 

 A vacuum of around $6 \times 10^{-6}$ mbar is
 achieved through a combination of a roots and rotary vane rough pump set,
  and two ISO 500 standard cryopumps. 
 The plant can coat the mirrors either in Aluminium,
 or in protected silver for optimal infrared performance 
 (see Sect.~\ref{sec-encstat} for the status of silver coating). 

\subsection{Handling operations} 

As above, all major lifting operations use the main Enclosure
 roof crane, after first removing a sector of the observing floor
 also using the crane. 
  For M1 re-coating, a summary of the 
  procedure is as follows: 
 \begin{enumerate} 
 \item 
 The telescope is pinned at horizon-pointing, and the
  Camera and then M2 are removed and stored 
  at ground level.
 \item  
  The complete top-end structure (top ring, spiders and
  trusses)  is detached from the Altitude ring and secured to 
  a dedicated cradle on the observing floor. 
\item
  A pair of 4.5 tonne 
  counterweights are attached to the Altitude ring to restore 
  balance of the telescope tube; then
  the tube is hand-cranked to zenith pointing and pinned.  
\item 
 The M1 restraint clamp and lateral supports are removed; 
  and the M1 cover is assembled
 at ground level and lifted onto the M1.  
\item 
 The M1 with cover is lifted vertically from the Cell 
  using a lifting plug passing through the Cassegrain rotator,
  then lowered down to the wash-stand on the ground floor. 
\item 
 The cover is lifted back up and stowed in the Cell. 
 The M1 coating is chemically stripped, and the mirror is then
  rinsed. 
\item 
 The M1 is lifted clear and the wash stand is removed. 
  Then the coating vessel lower section 
 drives underneath the mirror, and the M1 is lowered into the
  vessel.  
\item 
 The coating vessel lower section carries M1 back to the coating area,
 and the upper section closes for the coating process.  
\end{enumerate}  
Re-fitting is the reverse of the above. 

 Though there are numerous steps, 
  the M1 does not leave the building, and the main crane does 
 all the lifting. 

\subsection{Enclosure performance} 
\label{sec-encstat} 

 Overall the performance of the enclosure has been good: 
 the dome rotation is very smooth and quiet, the rotation 
 and slit doors have never had a serious mechanical jam to date, 
  and only very minor water leaks have occurred {\newtwo (now believed
  fixed)}.  The design 
 provides excellent ventilation and good protection from windshake; 
 the air-cooling system works well and has
 ample power for all weather conditions, although it is rather noisy.  

 Moderate problems were encountered with electrical and control issues:
 in particular, the early system had all-independent speed servos on the
 four rotation motors; inconsistencies between tacho readings could 
 lead to motors ``fighting'' each other
 then tripping out on excess current. A medium-term fix was to install
  larger brake resistors which made the fault rare, 
   but eventually the control system was modified to a master-slave 
 arrangement. 

 It turned out that the original 
 auxiliary building was somewhat too small for two reasons:   
 the original mirror wash area in the enclosure base (below the 
  lifting hatch) 
  is a busy area which could not be kept optimally clean, and also 
  storage space was very limited so the ground floor became cluttered 
  by various items of handling equipment. 
 For these reasons, in 2013--14 a 
 $9 \times 10$m extension was added on the far end
 of the auxiliary building, and the coating plant
 was shifted away from the telescope.  This provides a dedicated
 clean mirror wash area inside the auxiliary building, and a new 
  storeroom.  

 The initial silver coating with NiCr protective layer gave
 excellent reflectivity but was not quite durable enough, showing
   significant degradation
 after $\sim 1$ year on M1; thus, in April 2011 both mirrors were re-coated 
  in Aluminium. As of 2014, the 
 coating plant is being upgraded with a 4th magnetron to enable a
 Silicon Nitride protective overcoat based on the Gemini process, 
 and it is hoped this should produce a long-life silver coating 
  in the near future.  

{\newtwo Total average power consumption is 103\,kW (Weilenmann et al 
  \cite{weilenmann}), 
 including substantial contributions from the chiller (35\,kW) and
  Helium compressors (20 \,kW).}

\section{Software and electronics} 
\label{sec-soft} 

\subsection{Telescope control software and electronics} 

The VISTA telescope control software 
 mainly re-uses the ESO VLT control 
 software, giving a `look and feel' similar to the VLT control
 screens, so that Paranal telescope operators may readily swap 
 between VLTs and VISTA. 
 This software is largely written in C++ and runs on standard Linux PCs. 
 Special VISTA-specific modules were written
  for the active optics and enclosure control; significant
  modifications were made to the Preset module, 
 and minor modifications were made to the other modules where necessary 
 (Terrett \& Stewart \cite{terrett}).   

 The high-level control software does not control hardware
  directly, but sends commands to 
 Local Control Units (LCUs) which actually control the hardware. 
 The LCUs are small diskless computers running
 the VXWorks real-time operating system and programmed in C, 
 and there is one independent LCU per moving axis or subsystem.  
There are eleven LCUs in total: one for each Telescope axis, 
 one each for M1 supports and M2 hexapod, one for VIRCAM, 
 four for the two autoguiders and two low-order wavefront sensors, 
  and one for the Enclosure control.   

 Customised electronics units are also used for the low-level
  control of the telescope axis servo loops;
  also a CANbus unit is used to control the M1 pneumatic supports, and
  a PMAC controller is used for the M2 Hexapod motor drives.  

\subsection{Observing queue} 

For the high-level observation control, 
 the standard ESO BOB (Broker for Observation
 Blocks) is used; this takes Observation Blocks prepared
 by Phase 2 Preparation Package (P2PP) 
 for Surveys (Bierwith et al \cite{p2pp}).  
 For tiling user-defined areas of sky,  
  a package called the Survey Area Definition Tool (SADT) was developed: 
 this automatically defines suitable overlapping pointings with 
  selected guide and active optics stars, and creates an XML file
  fed to P2PP which populates these into observing blocks. 
 This automation is important since each rectangular VISTA tile 
 includes six distinct telescope pointings (pawprints) to provide
 uniform filling-in of detector gaps, and each pawprint in turn requires 
  one guide star and two LOWFS stars.  
 For redundancy, the software selects a ranked list of up to five candidate 
  stars for each of the above, so up to $6 \times 3 \times 5 = 90$
   candidate guide/LOWFS 
   stars may be selected for a single tile.  By default the software
 automatically selects the top-ranked star for each sensor, 
 but the telescope operator can override this choice 
 in the event of unsuitable star(s).  

\section{Assembly and Commissioning} 
\label{sec-commiss} 
 The official project kickoff was in April 2000. 
 The Phase A design phase was closed out in September 2001, 
  and essentially all the major contracts were in place
  by early 2003.  During 2003--4, the mountain peak was flattened, the
  new branch road was built and asphalted, 
  and the enclosure concrete base was completed. 
  In parallel, during 
  2003--4 the final design reviews took place 
  for the various subsystems. 
 Generally, the progress during manufacture
  and assembly went as planned, though a number of subsystems
  arrived slightly later than planned. 
 The enclosure was weatherproof in mid-2005, and
 the telescope structure was installed on site and tested 
  throughout 2006.  The complete VIRCAM 
  arrived at Paranal in January 2007, but
 unfortunately at that point neither of the mirrors was completed.   

 A ``first glimmer'' was achieved in March 2007, using
 a small 20\,cm Maksutov telescope mounted to the dummy instrument
 mass on the Cassegrain
 rotator, observing the sky through a pre-designed hole through
  the top-end barrel structure.   This run proved useful
 to build a preliminary pointing model and tune the telescope control
  software, before arrival of the real mirrors.  

 The secondary mirror arrived on site in May 2007, followed by the
 primary mirror in March 2008. To save time, 
 the primary mirror was transported by air from Moscow
  to Antofagasta in a specially chartered Antonov cargo aircraft. 
  After this, an intensive
 period of on-sky commissioning and debugging followed. 
 In the initial phase, a small and simple test camera (Puntino) 
  with imager and Shack-Hartmann wavefront sensor was
  used, to build an initial pointing model and active optics lookup tables; 
 then the test camera was replaced with VIRCAM in June 2008, and VIRCAM
  was used through the following 14 months of commissioning.  
 During commissioning, we were pleased to find that there
  were no severe design flaws and the system was capable
  of good performance; however 
  the commissioning process took significantly longer
  than planned due to a rather large number of relatively
   mundane technical glitches (notably wiring/connection problems, 
   unstable power supply units and glycol supply problems), 
  a lot of software debugging (especially on wavefront sensing)  
   along with a few more serious snags:  
  the latter are outlined in more detail in 
  Emerson \& Sutherland (\cite{es10}). Science verification 
  occurred in October 2009, followed by
 formal handover from VISTA Consortium via STFC
  to ESO ownership in December 2009, as part of the UK in-kind 
  contribution to joining ESO.  

\section{Observing, data processing and archiving} 
\label{sec-data} 

Here we give a short overview of the high-level observation
 software, VISTA Data Flow System (VDFS; Emerson et al \cite{em04}) 
 and VISTA Science Archive (VSA); 
 this is only a brief outline and we refer to relevant papers
 for more details. 

\subsection{Jittering and pawprints} 
 The VIRCAM detectors sparse-fill
 the focal plane, with gaps of 0.9 detector width in the
 $x-$direction and 0.425 detector width in the $y-$direction. 
 For gap-free sky coverage, 
  a set of six offset pointings (known as {\em pawprints}) 
  gives one filled rectangular {\em tile}. One tile 
  consists of a central rectangle $1.475 \times 1.017$ degrees 
  covered by at least {\em two}
  of the six pawprints; plus two thin stripes each 0.092 deg wide 
  (along the two long edges) covered by one pawprint.  
 In practice, each pawprint is usually comprised of several or many 
  offset {\em jitter} positions with typically $\sim 15$ arcsec offsets,  
  for optimal removal of detector artefacts in later processing. 
 The standard 6-point pawprint pattern (in several permutations of order) 
  and various jitter patterns are defined as standard ESO 
  templates, for convenience. 
 For telescope jitter movements, the same guide and active optics stars must be 
  re-used before and after (moving the readout windows in software),  
  so the SADT software avoids stars falling too close to a detector edge. 
 Moves to a new pawprint always require new guide and AO stars. 
  Typically a jitter move takes $\approx 7$\,sec, 
  and a pawprint move $\approx 10$\,sec. 
 Much additional information is available in the ESO VIRCAM User Manual
 \footnote{
 The VIRCAM user manual is at \\ 
  {\tt www.eso.org/sci/facilities/paranal/ \\
  instruments/vircam.html } }.  

\subsection{Data processing} 
 Quick-look data processing (QC0) is carried out in real-time at
 Paranal (Hummel et al \cite{qc0}); 
 this is primarily for timely assessment of data quality and
  detection of problems, not a final reduction.  
 The data volume is $\sim 300\,$GB per night average; until 2012 this
 was sent back on USB disks by air-freight. Since 2013, Paranal has
 been connected by optical fibre to the main Chilean internet 
 (delivered by the EVALSO project, Lemke et al \cite{evalso}),
  and the data now transfers over the internet. 
 The standard data pipeline is run at the Cambridge Astronomical
 Survey Unit (CASU); full details are given in Lewis et al (\cite{vdfs}), 
 and we summarise the steps here. Substantial information
 is available at the CASU web pages.\footnote{
 The CASU-VDFS technical information is at \\
  {\tt http://casu.ast.cam.ac.uk/surveys-projects/ \\
 vista/technical }} 

 \begin{enumerate} 
 \item 
 Reset correction happens automatically in the double-correlated sampling,
 so is not specifically a processing step. 
 \item Dark correction is done using dark frames. 
 \item Linearity correction is done using polynomials, fitted to 
  sequences of dome-flat frames. 
 \item Flat-field correction is done using twilight sky flats. 
 \item Sky background correction is done using various median operations
 on object-masked frames. 
 \item ``Destriping'' corrects an effect of horizontal non-repeatable
   stripes which are common across groups of four detectors, hence
 originates in the IRACE electronics.  
 \item Stacking: at this point the set of jittered images for
  a single pawprint are shifted and combined into a single stacked image
  with bad-pixel rejection. 
\end{enumerate} 
 After this, catalogue generation is run and the catalogues are
 astrometrically and photometrically calibrated using 2MASS stars. 

\subsection{Archiving and access} 
Processed  data is archived both at VISTA Science Archive (VSA), 
 Edinburgh, and at ESO.  The VSA carries out many additional processing steps
 including combining processed pawprints into tiles, 
 associating catalogue detections between different passbands, 
 and constructing matching tables against external catalogues
 such as SDSS and 2MASS. 
 The VSA user access includes a sophisticated SQL Server engine 
  with numerous indexed variables, 
 enabling fast processing of advanced queries. 
 Many details of the VSA system are provided by Cross et al (\cite{vsa}).

\section{System performance and public surveys} 
\label{sec-perf} 

\subsection{Performance summary} 
Generally, we are pleased to report that the overall system  
 works very well; 
 the VISTA telescope with VIRCAM routinely delivers 
 excellent wide-field images with sensitivity exceeding the
 original specifications (due notably to the high detector QE).   

The system has been in routine operation since
 November 2009, and the median delivered image quality 
  is $\approx 0.9$ arcsec, with a slight trend 
 with wavelength. 
 The 10th-percentile values are around 0.7 arcsec, 
  and images below 0.6 arcsec FWHM averaged across
 the full field are not uncommon. 

The system reliability has improved as expected over time as 
 glitches are fixed and/or workarounds are implemented to minimise
 impact. Technical time loss was around 10\% during the first 
 year, but has steadily improved and is now comparable to the
 VLT's; this is good considering that VISTA as a  
 single-instrument telescope is more vulnerable to instrument faults.  

Typical measured system zeropoints (for 1 ADU/sec) are
 given in Table~\ref{tab-zeropt}. 

\begin{table}
\caption{Typical VISTA system zeropoints, for broadband filters} 
\label{tab-zeropt} 
\centering 
\begin{tabular}{l c c c c c} 
\hline 
Filter &  Zeropoint  &   Zeropoint   &   AB-Vega offset  \\
       &  (Vega, 1 ADU/s) &  (Vega, $1e^-$/s) &   (mag)   \\
\hline 
 Z  & 23.95  & 25.51  & 0.52  \\ 
 Y  & 23.50  & 25.06  & 0.62  \\ 
 J  & 23.79  & 25.35  & 0.94  \\ 
 H  & 23.89  & 25.45  & 1.38 \\ 
 $\Ks$  & 23.06  & 24.62  & 1.84  \\ 
\hline
\end{tabular} 
\end{table}  

\subsection{Astrometry}   

The astrometric accuracy of VIRCAM images is good; 
 mean residuals from the fifth-order distortion pattern
 are below 0.025 arcsec.  Comparison of overlaps between 
 different tiles indicates typical systematic (absolute) offsets 
 $\sim 0.05$ arcsec at high galactic latitude; 
 this is probably limited by random
 noise (in 2MASS) on the $\sim 50$ 2MASS stars used per pawprint, but
 is already comfortably good enough for object matching
  to other wavebands and followup spectroscopy.   
 Differential astrometry within one tile is likely to be substantially
 better than this for sufficiently bright 
 objects.  Further calibration with e.g. the early-release {\em Gaia}
 catalogues will reveal the ultimate astrometric accuracy
 achievable with VISTA.   

\subsection{Photometry} 

 Briefly, VISTA photometry in the archive is given in the native VISTA 
 filter system (relative to Vega); this is quite similar
 to the WFCAM system except for VISTA's $\Ks$ filter. 
 The actual zeropoint for each pawprint is determined 
  using matching 2MASS stars. 
 Firstly, a large set of images on stable nights is used to determine 
  mean colour equations between 2MASS and VISTA systems, 
  and these (fixed) equations are then used
 to convert actual 2MASS magnitudes to predicted VISTA-system magnitudes
  for the 2MASS stars in a given pawprint.  The median difference between
  predicted and instrumental VISTA magnitude (after 
 distortion correction) is then used to set the
 oveall zeropoint for a VISTA pawprint.   This procedure
  is good for J,H,$\Ks$, but somewhat more uncertain at Z,Y where
  extrapolation from 2MASS is required. 
  Additional corrections
  (known as ``grouting'') are required for a catalogue derived
  from a tile.  

\subsection{Image artefacts} 

 The VIRCAM raw frames contain $\sim 1$ percent dead or
 hot pixels, but these are removed very well by jittering and 
  image stacking.  There is a problem with one half of 
 Detector 16 which exhibits unstable flat-fields,
 particularly at ZYJ bands, and is generally rejected for science
 data. 
  Detector persistence is much lower than 
 UKIRT WFCAM, but not entirely negligible for very bright stars. 
 No cross-talk between readout channels has been observed. 
 The most prominent artefacts are the filter ghosts around
 very bright stars, and also some smaller ``jets'' near 
 moderately bright stars.   The jets run perpendicular
 to detector edges, and appear to arise when 
 a star lands a few arcsec {\em outside} a detector edge; these  
 are suspected to be an internal reflection from structure at the detector
  edge.  These ghosts and jets are generally easy to recognise due
  to the proximity of the parent star. 

\subsection{Example images} 

The VISTA images are very large (16k by 12k pixels for a filled tile image) 
 so reproduction on A4 paper or computer screen 
  results in drastic loss of detail. 
Online zoomable images are much better to preserve information content,
 and a selection of these are available at \\
 {\tt http://www.eso.org/public/images/ \\
 archive/zoomable/?search=vista } 

\subsection{Public Surveys} 
\label{sec-survs} 

For the first five years of operations,  
  over 75\% of the VISTA observing time
 is being allocated to six large Public Surveys; these
 were selected by a dedicated ESO panel with a view
  both to standalone scientific merit, and also 
   wide-ranging legacy value of the data to the general community.  
 The remaining time is allocated to smaller
 PI-style programs via the standard ESO proposal process. 

The six surveys comprise one Hemisphere survey; 
 one Galactic bulge/plane survey; one Magellanic Cloud survey;
 and three nested extragalactic surveys with a range of area and
 depth.  Here we just provide a short summary of each survey,
 with a reference to more details. 

\begin{itemize} 
\item VHS (VISTA Hemisphere Survey; McMahon et al \cite{vhs}):  
 this covers almost the full Southern Hemisphere, $\delta < 0$, 
 in at least J and $\Ks$ passbands.  In more detail, there are three
 strategies for different sub-regions. The DES survey
 area will be covered with 120 sec exposures in J,H,$\Ks$ bands. 
 The VST-ATLAS area will be covered with 60 sec exposures in 
 the four bands Y,J,H,$\Ks$. The remaining area (about half)
  will be covered with 60 sec exposures at J,$\Ks$ only.  
 This survey is providing the core selection for the {\em Gaia}-ESO
 spectroscopic survey, and will provide important near-IR photometry
 to complement the GAIA mission.  

\item VVV (VISTA Variables in Via Lactea; Saito et al \cite{vvv-dr1}). 
  This survey covers
  a total of $500 \sqdeg$ in the Galactic bulge and inner plane,
 with substantial multi-epoch coverage.  The complete area 
 was covered in all five broadbands in the 2010 season, and subsequent
 seasons concentrate on multi-epoch sampling, with up to 100
 epochs mainly in $\Ks$ band. 

\item VMC (VISTA Magellanic Cloud survey; Cioni et al \cite{vmc}). 
  This survey covers both Magellanic Clouds and part of the Bridge,
  mainly in Y,J,$\Ks$ bands. The survey depth is intended 
 to reach the main-sequence turnoff in the $\Ks$ band, except in
 the most crowded fields. Around 12 epochs of $\Ks$ band are
 obtained for variability information. 

\item VIKING (VISTA Kilo-degree Infrared Galaxy Survey; 
 Edge et al \cite{edge}).  
  This is a medium-depth extragalactic survey 
 of $1500 \sqdeg$, approximating the 2dFGRS areas, 
 with coverage matching the VST-KIDS visible survey (de Jong et al 
  \cite{dejong}).  Coverage includes 
 all the {\em Herschel}-ATLAS survey fields except the Northern field, 
 and the GAMA redshift survey fields.  
 An early science highlight is discovery of three quasars 
  at $z = 6.6,\; 6.8, \; 6.9$ respectively (Venemans et al 
 \cite{venemans}), using NTT $i,z$ followup and
  and FORS spectroscopy. 
 (These are currently the second, third and fourth highest
  redshift known quasars).  

\item VIDEO (VISTA Deep Extragalactic Observations; Jarvis et al 
   \cite{video}). 
  This is a deep survey covering $12 \sqdeg$, in three well-studied
 extragalactic survey fields (extended {\em Chandra} Deep Field South; 
  XMM-LSS; and ELAIS-S1), in all five VISTA broadband filters. 
 Exposure times per pixel are 4-8 hours per passband. 
  All fields have existing mid-IR coverage from the {\em Spitzer}
  SWIRE and SERVS projects,
  and sub-mm coverage from {\em Herschel} HERMES, and are likely to be prime
   targets for future SKA pathfinders and ALMA.   
 The science goals include the galaxy population at $1 < z < 2.5$,
   high redshift clusters; low-luminosity AGNs; and identification
 and photometric redshifts of submm sources from the {\em Herschel} 
 HERMES surveys. 

\item UltraVISTA (Ultradeep VISTA survey; 
   McCracken et al \cite{ultravista}). 
  This is an ultra-deep survey covering a single VISTA tile, covering
  most of the HST-COSMOS field.  Along with the UKIDSS-UDS field, this is the
 best-studied degree-sized region in the extragalactic sky, and has a wealth
  of multiwavelength data from most major ground 
  and space observatories. 
 The first year's observations in 2010
  delivered a significant
  advance on previous CFHT-WIRCAM imaging sensitivity, 
  and added the valuable Y passband.  
  Further seasons are providing substantially deeper data  
    on half of the field (covering four disjoint stripes).  
 A recent highlight is discovery of 4--10 good candidate $z \approx 7$
  galaxies, the brightest examples known (Bowler et al \cite{bowler}).  
\end{itemize} 

%

\subsection{Longer term and 4MOST} 

The projected completion dates for the above VISTA public surveys 
 are around early 2017 (subject to continued approval). 
For the longer term future,  the ESA {\em Euclid} space mission
 is approved for launch around 2020.  The {\em Euclid} 
 field of view and near-IR pixel
 count are similar to VISTA, but given the much lower near-IR sky foreground
 in space, {\em Euclid} will deliver substantially fainter sensitivity 
 relative to VISTA,  by $\sim 2 - 3$ magnitudes.  
However, {\em Euclid} has several restrictions including: no $\Ks$ band;  
  very limited planned coverage at low galactic latitudes 
 (unless there is a substantial mission extension);  
 lack of narrowband filter capability (quasi-Y, J, H filters only); 
 and limited monitoring capability away from the
  ecliptic poles (due to Sun-angle constraints).   
 VISTA does not have these restrictions and thus offers significant
  complementarity.   

 Although spectroscopy did not form 
 part of the original VISTA project objectives, the telescope's 
 combination  of very wide field and large instrument capability 
  at Cassegrain focus turns out to be very suitable for a future 
  large-multiplex fibre spectrograph.  
 A second-generation VISTA instrument 
  known as 4MOST (4-m Multi-Object Spectroscopic Telescope,  
  de Jong et al \cite{f-most}) has been approved by ESO and is
  entering the preliminary design phase.  
  The baseline specifications for 4MOST include a new 4-lens visible 
 corrector with ADC giving 2.5 degree diameter
  field of view; and a multiplex of 2400 individually
  positioned fibres using an Echidna-style spine positioner 
   (simultaneously 1600 fibres to two medium-resolution spectrographs,
  and 800 fibres to one high-resolution spectrograph). 
   The 4MOST instrument offers very exciting science
 into the {\em Euclid} era, both for followup of the GAIA and eROSITA
 space missions, large $z \simlt 1$ redshift surveys, 
  and followup of the wide-area imaging surveys
  (especially VHS, VIKING and VMC), with a predicted yield
 of $\sim 25$ million spectra in a 5-year merged science programme.    

 We also note that around 2.5 years of VIRCAM time
 will elapse between the projected completion of the public surveys above 
 and the delivery of 4MOST,
 so there are likely to be opportunities for
 novel large projects using VIRCAM in this timeframe.   

\section{Conclusions} 
\label{sec-conc} 

We have outlined the design, construction and performance
 of the 4.1m wide-field VISTA telescope at Cerro Paranal
  and its 67 Mpixel infrared camera (VIRCAM),   
 from the Phase A study in April 2000 to the start of science
  operations in October 2009 and early science results. 
  We have described several of
 the novel design features, including the $f/1.0$ primary,
  cold-baffled IR camera with dichroic baffle coating, 
 and the 5-axis closed-loop collimation system, using dual off-axis 
  curvature wavefront sensors. 

In the medium term, VISTA with VIRCAM is expected to remain the world's fastest
 wide-area near-infrared imaging system, and the only such 
 system in the Southern hemisphere,  until the predicted launch of
 the ESA {\em Euclid} space mission in the early 2020's.  
 The six ESO public surveys are making steady progress, 
 and many interesting science results are emerging, as briefly outlined
 above.  These surveys will be also very important resources for 
 target selection for upcoming major facilities including ALMA,
 SKA pathfinders, MOONS and JWST.  

(This is an author-produced version of this paper for arXiv.org: 
 the definitive version of record
 is published in A\&A, and available at Digital Object Identifier \\
 DOI:10.1051/0004-6361/201424973 ) 

\begin{acknowledgements} 
 The VISTA project was made possible by funding from the 
 UK Joint Infrastructure Fund (JIF) and PPARC (later STFC). 
 Overall systems design and management  
 was done by the VISTA Project Office at UKATC, Edinburgh, and the
 IR Camera was led by RAL Space with substantial contributions from
 UKATC and University of Durham.  

 The participating institutions were Queen Mary University of London
 (Lead institution); 
 Queen's University Belfast; Univ. of Birmingham; Univ. of Cambridge; 
 Cardiff University;  Univ. of Central Lancashire; 
 Univ. of Durham; Univ. of Edinburgh; 
 Univ. of Hertfordshire; Keele University; Univ. of Leicester; 
 Liverpool John Moores University; 
 Univ. of Nottingham;  Univ. of Oxford; 
 Univ. of St. Andrews;  Univ. of Southampton;  
 Univ. of Sussex; University College, London. 
  
 We are grateful to early advice from ESO staff including
 Stefano Stanghellini, Massimo Tarenghi and Martin Cullum.  
 {\newtwo Mark Casali}, 
 Sue Worswick and Martin Fisher made substantial contributions
 in the Phase A design studies, and the late Richard Bingham
  provided detailed design for the WFS/AG units.  

 Numerous individuals served on the VISTA Science Committee 
 (Chair: Andy Lawrence) which drafted the science requirements; 
  project oversight 
 was provided by the VISTA Executive Board (Chair: Mike Edmunds) 
 and later the VISTA Project Board (Chair: Richard Wade) with  
 Robert Laing and Pat Roche as external members.  
 
 We thank the many technical experts who assisted by 
 serving on the numerous review panels during the course
 of the project, and the many engineers at UKATC, RAL Space, 
  LZOS, Vertex-RSI, Raytheon Vision Systems, 
  NTE Barcelona and EIE Venice who contributed 
 major efforts in turning the project into reality.  

 We thank many ESO staff, especially Carlos La Fuente, 
 Andres Parraguez, Pascual Rojas, Thomas Szeifert and
 Valentin Ivanov for invaluable support during the commissioning 
 phase and thereafter. 

\end{acknowledgements}


\end{document}